\documentclass[11pt]{article}
\usepackage{amsmath,amsxtra,mathrsfs,amssymb}
\usepackage{upgreek}
\usepackage{natbib}

\usepackage[english]{babel}
\usepackage[T1]{fontenc}

\newcommand{\at}{{\char '100}}

\newcommand{\DEF}{\overset{\mathrm{def}}{=}}
\newcommand{\DEFt}{\smash{\overset{\text{\tiny def}}{=}}}

\newcommand{\T}[1]{#1^{\mathrm{\scriptscriptstyle T}}}
\newcommand{\Tphantom}[1]{#1^{\phantom{\mathrm{\scriptscriptstyle T}}}} 
\newcommand{\Tsmall}[1]{#1^{\mathrm{\tiny T}}} 
\newcommand{\Poisson}[2]{\left\{\smash{#1,#2}\right\}}
\renewcommand{\theenumi}{(\kern -0.15ex{\roman{enumi}})}

\newcommand\bra[1]{\mathinner{\langle{#1}|}}
 \newcommand\ket[1]{\mathinner{|{#1}\rangle}}
\newcommand\braket[2]{\mathinner{\langle{#1}|{#2}\rangle}}

\newcommand{\tv}[1]{\mathbf{#1}}

\newcommand{\imat}{\mathrm{i}}

\newcommand{\dmat}{\mathrm{d}}
\newcommand{\Omat}{\mathrm{O}}

\newcommand{\Dsuiv}{D} 
\newcommand{\grads}{\partial_{\tv{x}}}

\DeclareSymbolFontAlphabet{\mathpxbb}{AMSb}

\newcommand{\RR}{\mathpxbb{R}}

\newcommand{\EXP}[1]{\mathrm{e}^{#1}}
\newcommand{\finiteset}[2]{\left\{#1,\ldots\kern -0.15ex,#2\right\}}

\newcommand{\Dim}{{\mathchoice{\text{\footnotesize \rm D}}{\text{\footnotesize \rm D}}{\textsc{D}}{\textsc{d}}}}
\newcommand{\Deg}{{\mathchoice{\text{\footnotesize \rm L}}{\text{\footnotesize \rm L}}{\textsc{\rm L}}{\textsc{l}}}}

\begin{document}

\title{Applications of Noether conservation theorem to Hamiltonian systems}

\author{Amaury Mouchet\\ \small
Laboratoire de Math\'ematiques
  et de Physique Th\'eorique, \\ \small
Universit\'e Fran\c{c}ois Rabelais de Tours --- \textsc{\textsc{cnrs (umr 7350)}},\\ \small
 F\'ed\'eration Denis Poisson,\\ \small
 Parc de Grandmont 37200  Tours,  France\\ \small
 mouchet\at phys.univ-tours.fr\\ \small
\date{\today}}

\maketitle
\begin{abstract}
The Noether theorem connecting symmetries and conservation laws
can be applied directly in a Hamiltonian framework without using any
intermediate Lagrangian formulation. This requires a careful
discussion about the invariance of the boundary conditions under a
canonical transformation and this paper proposes to address this
issue.  Then, the unified treatment of Hamiltonian systems offered by
Noether's approach is illustrated on several examples, including
classical field theory and quantum dynamics.
\end{abstract}

\noindent\textsc{pacs:} 11.30.-j,      
04.20.Fy, 	
11.10.Ef, 	
02.30.Xx 	

\bigskip



\section{Introduction}\label{sec:introduction}

After its original publication in German in 1918, and even though it
was first motivated by theoretical physics issues in General
Relativity, it took a surprisingly long time for the physicists of the
twentieth century to become aware of the profoundness of Noether's
seminal article (see~\cite{KosmannSchwarzbach11a} for an English
translation and a historical analysis of its impact, see
also~\cite[\S\;7]{Kastrup83a} and~\cite{Byers94a}).
\nocite{Doncel+87a} Since then, about the 1950's say, as far as
theoretical physics is concerned, Noether's work spread widely from
research articles in more general textbooks and, nowadays, it even
reaches some online pages like Wikipedia's~\citep{wiki_noether}
intended to a (relatively) large audience including undergraduate
students (see also~\cite{Neuenschwander11a}
and~\cite[\S\;5.2]{KosmannSchwarzbach11b}).  However, the vast majority
of these later presentations, unfortunately following the steps
of~\citet{Hill51a} (see~\citet[\S\;4.7]{KosmannSchwarzbach11a}), reduces
drastically the scope of Noether's article\footnote{Obviously, the
  common fact that research articles are more quoted than read is all
  the more manifest for rich fundamental papers.}; (i)~because they
commonly refer to the first main theorem (``\emph{The} Noether
theorem'') without even mentioning that Noether's 1918 paper contains
more physically relevant material\footnote{\label{fn:2ndtheorem} There
  is a second main theorem establishing a one-to-one
  correspondence between Gauge invariance and some identities between
  the Euler-Lagrange equations and their derivatives (see \S\;\ref{subsec:2theorems} below).  These Noether
  identities render that a gauge-invariant model is necessarily
  a constrained Hamiltonian/Lagrangian system in Dirac's
  sense~\cite{Dirac64a}.  Furthermore, a by-product result also proven
  by Noether~\cite[\S\;5]{Noether18a} is that the constants of motion
  associated, through the first theorem, with an invariance under a
  Lie group are themselves invariant under the transformations
  representing this group.}  and also (ii)~because the connection
between the existence of a conservation law and some invariance under a
continuous group of transformations in a variational problem is
predominantly illustrated in a Lagrangian framework, for
instance~\cite[\S 7.3]{Weinberg95a}, (not to speak that the order of
the derivatives involved in the Lagrangian do not generally exceed
one, albeit Noether explicitly works with integrands of arbitrary
orders). As a consequence, an enormous literature flourished that
claimed to generalise Noether's results whereas it only generalised
the secondary poor man's versions of it without acknowledging that
these so-called generalisations were already present in Noether's
original work~\cite[\S\;5.5]{KosmannSchwarzbach11a} or in
Bessel-Hagen's paper~\cite{BesselHagen21a} --- directly owed to a ``an
oral communication from Emmy Noether'' (see also~\cite[\S\;4, footnote
  20]{Noether18a}) --- where invariance of the integrand defining the
functional is considered ``up to a divergence''.

Nevertheless, fortunately, the success of gauge models in quantum
field theory motivated several works where Noether's contribution was
employed in (almost) all its powerful generality (for articles not concerned
by~(i) see for instance~\cite{Barbashov/Nesterenko83a,Lusanna91a} and
the more epistemological approach proposed
in~\cite{Brading/Brown03a}).\nocite{Brading/Castellani03a} To
counterbalance~(ii), the present paper is an attempt to provide a
unified treatment of Noether's conservation laws in the Hamiltonian
framework, i.e. where the canonical formalism is used. In this
context, the advantages of the latter have already been emphasized by
a certain number of works among which we can
cite~\cite{Henneaux/Teitelboim92a,Li93a,Deriglazov/Evdokimov00a} where
the main focus was naturally put on the Noether's second theorem (see
footnote~\ref{fn:2ndtheorem}) but not necessarily, since classical
mechanics was also considered --- \cite{Sarlet/Cantrijn81a},
regrettably suffering of flaw~(i) --- even with pedagogical purposes
\cite{Leubner/Marte85a}, \cite[\S\;7.11]{Deriglazov10a}. The main
advantage of the Hamiltonian approach over the standard Lagrangian one
is that it incorporates more naturally a larger class of
transformations, namely the canonical transformations (in
phase-space), than the point transformations (in configuration
space). To recover the constants of motion associated with the
canonical transformations that cannot be reduced to some point
transformations, one has to consider some symmetry transformations of
the Lagrangian action that depend on the time derivative of the
degrees of freedom. Anyway, these so called ``dynamical'',
``accidental'' or ``hidden'' symmetries (the best known example being
the Laplace-Runge-Lenz vector for the two-body Coulombian model
\cite[\S\;5A]{LevyLeblond71b}) are completely covered by Noether's
original treatment, even if we stick to a Lagrangian framework.

As a starting point I will explain, in~\S\;\ref{sec:hamvarprinciple},
how the price to pay when working within the Hamiltonian framework is
that special care is required concerning the boundary conditions
imposed when formulating the variational principle: unlike what occurs
in the configuration space, in phase-space not all the initial and
final dynamical variables can be fixed arbitrarily but rather half of
them; the choice of which ones should be fixed is an essential part of
the model and therefore should be included in any discussion about its
invariance under a group of transformations.  As far as I know, in the
literature where Noether's work is considered, including
\cite{Noether18a} itself or even when a Hamiltonian perspective is
privileged, the invariance of the boundary conditions is not genuinely
considered and only the invariance of the functional upon which the
variational principle relies is examined. This may be understood
because as far as we keep in mind a Lagrangian formulation, the
boundary conditions are not generically constrained; on the other
hand, in a Hamiltonian formulation, there are some constraints that
fix half of the canonical variables and the invariance of the action
under a canonical transformation does not guarantee that the
constraints are themselves invariant under this transformation.  Since
the present paper intends to show how Noether's conservation laws can
be directly applied in a Hamiltonian context, I will have to clarify
this issue and for this purpose I propose to introduce
(\S\;\ref{subsec:boundaryfunction}) a boundary function, defined on
phase space, whose role is to encapsulate the boundary conditions 
and offers some latitude for changing them.  In
\S\;\ref{sec:invarianceconservation}, for a classical Hamiltonian
system I will present a derivation of the conservation laws from the invariance under the
most general canonical transformations. Then, before I show
in~\S\;\ref{subsec:finitedof} that the same results can be obtained
with Noether's approach, I will paraphrase Noether's original paper
in~\S\;\ref{sec:Noetherformulation} for the sake of self-containedness
and for defining the notations.  Before I briefly conclude, I will
show explicitly how Noether's method can be applied for models
involving classical fields (\S\;\ref{subsec:field}) and in quantum
theory (\S\;\ref{sec:quantum}). For completeness the connection with
the Lagrangian framework will be presented in
\S\;\ref{subsec:lagrangian}.

\section{Hamiltonian variational principle and the boundary conditions}\label{sec:hamvarprinciple}

\subsection{Formulation of the variational principle in a Hamiltonian context}

We shall work with a Hamiltonian system described by the independent
canonical variables~$(p,q)$ referring to a point in phase space.
Whenever required, we will explicitly label the degrees of freedom
by~$\alpha$ that may be a set of discrete indices, a subset of
continuous numbers or a mixture of both. For instance, for~$\Deg$
degrees of freedom, we
have~$(p,q)=(p_\alpha,q_\alpha)_{\alpha\in\finiteset{1}{\Deg}}$
whereas for a scalar field in a~$\Dim$-dimensional space we will
take~$\alpha=\tv{x}=(x^1,\dots,x^\Dim)=(x^i)_{i\in\finiteset{1}{\Dim}}$ and then~$(p,q)$ will
stand for the fields\linebreak $\big\{\pi(\tv{x}),\varphi(\tv{x})\big\}_{\tv{x}\in\RR^\Dim}$.  The
dynamics of the system is based on a variational principle i.e. it
corresponds to an evolution where the dynamical variables are
functions of time\footnote{We will never bother about the regularity of all
  the functions we will meet, assuming they are smooth enough for
  their derivative to be defined when necessary.} that extremalise
some functional~$S$ called the action.  In the standard presentation
of the Hamilton principle in phase space, see \cite{Ray73a} and its
references, the action is defined as the
functional~$\int_{t_i}^{t_f}\big(p\,\dmat q/\dmat t-H(p,q,t)\big)\dmat
t$ of the smooth functions of time~$t\mapsto\big(p(t),q(t)\big)$ (the
summation/integral on the degrees of freedom labelled by~$\alpha$ is left implicit).  When the
Hamiltonian~$H(p,q,t)$ depends explicitly on time~$t$, it is often
convenient to work in an extended phase space where~$(-H,t)$ can be
seen as an additional pair of canonical dynamical variables; we shall
not use this possibility but still, we shall keep some trace of the
similarity between~$q$ and~$t$ on one hand and between~$p$ and~$-H$ 
on the other hand
 by considering the action
\begin{equation}
  \label{def:actionhamiltonienne}
        S_0[p(\cdot),q(\cdot),t(\cdot)]\ \DEF\ 
         \int_{s_i}^{s_f}\left( p(s)\frac{\dmat q}{\dmat s}(s)-H\big(p(s),q(s),t(s)\big)\frac{\dmat t}{\dmat s}(s)
                         \right)
         \dmat s
\end{equation}
as a functional of~$s\mapsto p(s)$, $s\mapsto q(s)$ and $s\mapsto
t(s)$ where~$s$ is a one-dimensional real parametrisation. An
infinitesimal variation~$p(s)+\delta p(s)$, $q(s)+\delta q(s)$,
$t(s)+\delta t(s)$ induces the variation~$S_0+\delta S_0$ of the value of
the action where, to first order in~$(\delta p,\delta q,\delta t)$, we
have, with the customary use of integration by parts,
\begin{equation}\label{eq:deltaS}
\begin{split}
        \delta S_0&= p(s_f)\delta q(s_f)-p(s_i)\delta q(s_i)\\
               &-H\big(p(s_f),q(s_f),t(s_f)\big)\delta t(s_f)+H\big(p(s_i),q(s_i),t(s_i)\big)\delta t(s_i)\\
     &+ \int_{s_i}^{s_f}\left\{
    \left[\frac{\dmat q}{\dmat s}(s)-\partial_{p} H\big(p(s),q(s),t(s)\big)\frac{\dmat t}{\dmat s}(s)
   \right]\delta p(s)  \right.  
   \\
   & +\left[-\frac{\dmat p}{\dmat s}(s)-\partial_{q} H\big(p(s),q(s),t(s) \big)\frac{\dmat t}{\dmat s}(s)
    \right]\delta q(s)\ 
   \\
   &+\left.\left[\frac{\dmat}{\dmat s}H\big(p(s),q(s),t(s)\big)-\partial_{t} H\big(p(s),q(s),t(s)\big)\frac{\dmat t}{\dmat s}(s)\right]\delta t(s)
      \right\}\dmat s
\end{split}
\end{equation}
and, then, the Hamilton variational principle can be formulated as
follows: in the set of all phase-space paths connecting the initial
position~$q(s_i)=q_i$ at~$t(s_i)=t_i$ to the final
position~$q(s_f)=q_f$ at $t(s_f)=t_f$ the dynamics of the
system follows one for which~$S_0$ is stationary\footnote{This classical
  path is not necessarily unique and may be even a degenerate critical
  path for~$S_0$, see however the next footnote. }; in other words, the variation~$\delta S_0$ 
 vanishes in first
order provided we restrict the variations to those such that
\begin{subequations}\label{eq:condlim_qqtt}
\begin{equation}\label{eq:condlim_qq}
        \delta q(s_f)=\delta q(s_i)=0\;;
\end{equation}
\begin{equation}\label{eq:condlim_tt}
        \delta t(s_f)=\delta t(s_i)=0\;
\end{equation}
\end{subequations}
whereas the other variations~$\delta t(s)$, $\delta p(s) $ and $\delta q(s)$ remain
arbitrary (but small), hence independent one
    from the other.  Hamilton's equations
\begin{subequations}\label{eq:Ham}
        \begin{eqnarray}
        \frac{\dmat p}{\dmat t}&=&-\partial_{q}H(p,q,t)\;;\label{eq:Ham_pdot}\\
        \frac{\dmat q}{\dmat t}&=&\phantom{-}\partial_{p}H(p,q,t)\;;\label{eq:Ham_qdot}
        \end{eqnarray}
\end{subequations}
come from the cancellation of the two first brackets in the integrand
of~\eqref{eq:deltaS}, then the cancellation of the third one
follows. The restrictions~\eqref{eq:condlim_qqtt} on the otherwise
arbitrary variations~$\delta p(s)$, $\delta q(s)$, $\delta t(s)$ provides
sufficient conditions to cancel the boundary terms given by the two first
lines of the right-hand side of~\eqref{eq:deltaS} but they are not necessary, one could
impose~$\delta q$ to be transversal to~$p$ both at~$t_i$ and~$t_f$, or impose 
some periodic conditions (see footnote~\ref{fn:transversality}).

\subsection{The differences concerning the boundary conditions between Lagrangian and Hamiltonian models}

In the usual Lagrangian 
approach, the $q$'s constitute all the
dynamical variables and a generic choice of~$(q_i,q_f,t_i,t_f)$ leads
to a well-defined variational problem having one isolated
solution\footnote{In the space of initial conditions, the
  singularities corresponding to bifurcation points, caustics,
  etc. are submanifolds of strictly lower dimension (higher
  co-dimension) and therefore outside the scope, by definition, of
  what is meant by ``generic''. In other words we consider as generic any 
property that is structurally stable, that is, unchanged under
a small enough arbitrary transformation.}: no constraint on~$(q_i,q_f,t_i,t_f)$
is required and it is commonly assumed that the variations of all the
dynamical variables vanish at the boundary; any point
transformation~$q\mapsto\T{q}(q)$ preserves this condition since then
$\delta\T{q}=(\partial_q\T{q})\delta q$ and we have $\delta
q=0\Leftrightarrow\delta\T{q}=0$.

In a Hamiltonian framework, obviously, because the dynamical
variables~$q$ and~$p$ are not treated on the same footing in the
definition~\eqref{def:actionhamiltonienne} of~$S_0$, there is an
imbalance in the boundary conditions and in their variations
between~$\delta q$ and~$\delta p$. More physically, this comes from
the fact that the classical orbits, defined to be the solutions
of~\eqref{eq:Ham}, are generically determined by \emph{half} of the
set~$(p_i,q_i,p_f,q_f)$; in general, there will be no classical
solution for a given a priori set~$(p_i,q_i,p_f,q_f)$ and a
well-defined variational principle --- that is, neither overdetermined
nor underdetermined --- requires some constraints that make half of
these dynamical variables to be functions of half the independent
other ones.  Any canonical transformation, which usually shuffles
the~$(p,q)$'s, will not only affect the functional~$S_0$ but also the
boundary conditions required by the statement of variational
principle. For a canonical transformation the transformed dynamical
variables~$\T{q}$ and~$\T{p}$ are expected to be functions of both~$q$
and~$p$ and, then, as noted in \cite{Quade79a}, the 
conditions~\eqref{eq:condlim_qq} alone do not
imply that~$\delta\T{q}_i=\delta\T{q}_f=0$ since
neither~$\delta\T{p}_i$ nor $\delta\T{p}_f$ vanish in general.

In any case, the behaviour of the initial conditions under a
transformation should be included when studying the invariance of a
variational model but this issue is made more imperative 
in a Hamiltonian than in
a Lagrangian viewpoint, at least when the Lagrangian does not depend on time derivative of  order higher than one\footnote{\label{fn:addednotev2}Aside from the crucial question of instabilities (Ostrogradsky's, ghosts etc), even for higher orders, the correspondence between Lagrangian and Hamiltonian formalism can actually be formalised \cite[ \S~7.1, for instance]{Gitman/Tyutin90a} but in that cases
the boundary conditions are not so simple, even in the Lagrangian framework, as one can be easily convinced by considering~$L(q,\dot{q},\ddot{q})$. The attempts of building a path-integral formulation compatible
with General Relativity require a special care of boundary conditions even at the Lagrangian level:
this is of course the core of holography and of the AdS/CFT correspondence; 
 it has also been  noted by \cite{York72a}, \cite{Gibbons/Hawking77a} that a boundary term should be added 
to the Einstein-Hilbert action, in particular if one wants to keep the concatenation property 
(more on this in the next section). Section~\ref{subsec:lagrangian} suggests 
that, even in the simplest models, 
the relationship between the boundary term in the Hamiltonian formalism and the Lagrangian formalism
is not straightforward.
}.

\subsection{The boundary function}\label{subsec:boundaryfunction}

To restore some sort of equal treatment between the~$q$'s and the~$p$'s
in the Hamiltonian framework, one can tentatively add to~$S_0$ a function
$A$ of the dynamical variables at the end points~$(q_f,p_f,t_f;q_i,p_i,t_i)$ whose variations~$\delta A$ depend a priori on 
the variations of all the dynamical variables at the boundaries. Nevertheless
we will restrict the choice of~$A(q_f,p_f,t_f;q_i,p_i,t_i)$ to functions of the form~$B(q_f,p_f,t_f)-B(q_i,p_i,t_i)$
in order to preserve the \textit{concatenation property} according to which 
the value of the action of two concatenated paths is the sum of the actions of each of the two paths.
This strategy is equivalent to add to the integrand of~$S_0$ the total derivative 
of the boundary function~$B$ (see \cite[\S\;IV.5.1, footnote~1 p.~211]{Courant/Hilbert53a}):
\begin{equation}\label{def:actionhamiltonienneB}
        S_B[p(\cdot),q(\cdot),t(\cdot)]\ \DEF\ 
         \int_{s_i}^{s_f}\left( p\,\frac{\dmat q}{\dmat s}
         +\Big(-H(p,q,t)+\frac{\dmat }{\dmat t}B(p,q,t)\Big)\frac{\dmat t}{\dmat s}
                         \right)
         \dmat s\;.
\end{equation}
This modification does not alter Hamilton's
equations~\eqref{eq:Ham}\footnote{The fact that a total derivative can
  be added to a Lagrangian without changing the evolution equations is
  well-known for a long-time. As already noticed above it is mentioned
  by Noether \cite[\S\;4, footnote 20]{Noether18a} and this flexibility
  has been used for many purposes ; in particular in Bessel-Hagen's paper~\cite[\S\;1]{BesselHagen21a}, see also the discussion in \cite[\S\;3]{Brading/Brown03a}.  } but
allows to reformulate the variational problem within the set of
phase-space paths defined by the boundary conditions such that
\begin{equation}\label{eq:condlim_B}
        \Big[p\delta q-H\delta t+\delta B\Big]^{s_f}_{s_i}=0\;.
\end{equation}
For instance by choosing~$B(p,q,t)=-pq$, the roles of the~$p$'s and
the~$q$'s are exchanged and~\eqref{eq:condlim_qq} is replaced
by~$\delta p(s_f)=\delta p(s_i)=0$ whereas if we take~$B(p,q,t)=-pq/2$
the symmetry between~$p$ and~$q$ is (almost) obtained.

We see that the boundary function is defined up to a 
function of time only since the substitution
\begin{equation}\label{eq:uptob}
  B'(p,q,t)\DEF B(p,q,t)+b(t)\;;\qquad H'(p,q,t)\DEF H(p,q,t)+\frac{\dmat b}{\dmat t}(t)
\end{equation}
leaves unchanged both the action~\eqref{def:actionhamiltonienneB} and
the boundary conditions~\eqref{eq:condlim_B}. A dependence of~$b$ on
the other dynamical variables is unacceptable since it would introduce
time derivatives of~$p$ and~$q$ in the Hamiltonian.

\section{Transformation, invariance and conservation laws}\label{sec:invarianceconservation}

\subsection{Canonical transformation of the action, the Hamiltonian and the boundary function}

In the present paper we refrain to use the whole concepts and
formalism of symplectic geometry that has been developed for dynamical
systems and prefer to keep a ``physicist touch'' without referring to
fiber bundles, jets, etc. even though the latter allow to work with a
completely coordinate-free formulation. With this line of thought, we
follow a path closer to Noether's original formulation. However,
keeping a geometrical interpretation in mind, if we consider the
action~\eqref{def:actionhamiltonienneB} as a scalar functional of a
geometrical path in phase space, any canonical
transformation~$(q,p,t)\mapsto(\T{q},\T{p},\T{t})$ can be seen as a
change of coordinate patch (the so-called passive transformation on
which the geometrical concept of manifold relies) that does not affect
the \emph{value} of the action for the considered path, so we should
have
\begin{equation}\label{def:Stransforme}
         \T{S}_{\smash{\T{B}}}[\T{p}(\cdot),\T{q}(\cdot),\T{t}(\cdot)]
         \ \DEF  S_B[p(\cdot),q(\cdot),t(\cdot)]\;;
\end{equation}
in this point of view, the latter relation is a \textit{definition} of
the transformed functional, not an expression of the invariance of the
model.  The canonical character of the transformation guarantees
that~$\T{S}_{\smash{\T{B}}}$ takes the same form
as~\eqref{def:actionhamiltonienneB}, namely
 \begin{multline}
        \T{S}_{\smash{\T{B}}}[\T{p}(\cdot),\T{q}(\cdot),\T{t}(\cdot)]
         =  
     \int_{s_i}^{s_f}  \left( \T{p}(s)\frac{\dmat \T{q}}{\dmat s}(s)
         -\T{H}\big(\T{p}(s),\T{q}(s),\T{t}(s)\big)\frac{\dmat\T{t}}{\dmat \Tphantom{s}}(s)\right. \\
          \left. +\frac{\dmat}{\dmat s}
            \Big[\T{B}\big(\T{p}(s),\T{q}(s),\T{t}(s)\big)\Big] \right)
         \dmat s\;,
\end{multline}
which leads to a \textit{definition} of~$\T{H}$ and~$\T{B}$ up to a
function of time only (see~\eqref{eq:uptob}).  Since the
equality~\eqref{def:Stransforme} holds for any phase-space path
(whether classical or not), a necessary (and sufficient) condition is
that
\begin{multline}\label{eq:Htransfocanoniquegenerale}
        \T{p}\,\dmat \T{q}-\T{H}(\T{p},\T{q},\T{t})\,\dmat\T{t}+\dmat \big(\T{B}(\T{p},\T{q},\T{t})\big)
        \\=p\,\dmat q-H(p,q,t)\,\dmat t+\dmat \big(B(p,q,t)\big)\;,
\end{multline}
which provides an explicit expression for~$\T{H}(\T{p},\T{q},\T{t})$
and $\T{B}(\T{p},\T{q},\T{t})$ according to the choice of the
independent coordinates in phase-space.  For instance, if we pick up
$\T{p}$, $q$ and~$t$
 and assume that the transformation of time is given by a general function~$\T{t}(\T{p},q,t)$\footnote{
A notable case where~$\T{t}$ depends on~$q$ is provided by the Lorentz transformations.},
 the expression~\eqref{eq:Htransfocanoniquegenerale} in terms of the
corresponding differential forms is
\begin{multline}
        \T{q}\,\dmat \T{p}+p\,\dmat q+\T{H}(\T{p},\T{q},\T{t})\,\dmat\T{t}-H(p,q,t)\,\dmat t\\
        =\dmat \big(\T{p}\T{q}+\T{B}(\T{p},\T{q},\T{t})-B(p,q,t)\big)=\dmat F\;,
\end{multline}
which is the differential of a generating function~$F(\T{p},q,t)$ of
the canonical transformation implicitly defined (up to a function of
time only) by
\begin{subequations}\label{eq:Fpq}
\begin{eqnarray}   
      p&=&\frac{\partial F}{\partial q}-\T{H}(\T{p},\T{q},\T{t})\frac{\partial \T{t}}{\partial \Tphantom{q}}\;;  \\
      \T{q}&=&\frac{\partial F}{\partial \T{p}}-\T{H}(\T{p},\T{q},\T{t})\frac{\partial \T{t}}{\partial \T{p}}\;.
\end{eqnarray}
\end{subequations}
Then, we get
 \begin{equation}\label{eq:Htransforme}
        \T{H}(\T{p},\T{q},\T{t})\,\frac{\partial\T{t}}{\partial \Tphantom{t}}(\T{p},q,t)
         =H(p,q,t)+\frac{\partial F}{\partial t}(\T{p},q,t)
\end{equation}
and
\begin{equation}\label{eq:Btransforme}
        \T{B}(\T{p},\T{q},\T{t})=B(p,q,t)-\T{p}\T{q}+F(\T{p},q,t)\;.
\end{equation}
The substitution~\eqref{eq:uptob} corresponds to the alternative
choice~$F'\DEF F-b$.  From the latter relation, we understand why a
boundary function~$B$ has to be introduced in the definition of the
action when discussing the effects of a general canonical
transformation. Even if we start with a~$B$ that vanishes identically,
a canonical transformation turns~$B\equiv0$
into~$-\smash{\breve{F}}\big(\T{q},q\big)$ where~$\smash{\breve{F}}$
is the generating function given by the following Legendre transform
of~$F$
\begin{equation}\label{def:Fqq}
         \breve{F}(\T{q},q)\DEF \T{p}\T{q}-F(\T{p},q)\;,
\end{equation}
and therefore~$\T{B}\not\equiv0$ in general (this special case is the
point raised in \cite{Quade79a}).  In the particular case of point
transformations~$\T{q}=f(q,t)$, the boundary function can remain
unchanged since we can always choose \linebreak$F(\T{p},q,t)=\T{p}f(q,t)$ for
which~$\smash{\breve{F}}\equiv0$.

\subsection{What is meant by invariance}\label{subsec:clinvariance}

When talking about the invariance of a Hamiltonian model under a
transformation, one may imply (at least) three non-equivalent
conditions: the invariance of the form of the
action~\eqref{def:actionhamiltonienneB}, the invariance of the form of
Hamilton's equations~\eqref{eq:Ham} or the invariance of the form of
Newton equations derived from the latter\footnote{Reference \cite{Havas73a} provides illuminating examples
of the differences between
this three types of invariances.}.  As far as only classical
dynamics is concerned, the invariance of the action appears to be a
too strong condition: if only the critical points of a function(nal)
are relevant, there is no need to impose the invariance of the
function(nal) itself outside some neighbourhood of its critical points
and, provided no bifurcation occurs, one may substantially transform
the function(nal) without impacting the location and the properties of
its critical points. For instance the
transformation~$S\mapsto\T{S}=S+\epsilon\;\mathrm{sinh}S$, with
$\epsilon$ being a real parameter, would actually lead to the same
critical points\footnote{It is also easy to construct an example for
  which not only the critical points are preserved but also their
  stability as well as the higher orders of the functional derivatives
  of~$S$ evaluated on the classical solutions. }. However, by
considering that quantum theory is a more fundamental theory than the
classical one, from its formulation in terms of path integrals due to
Feynman\footnote{The original Feynman's formulation has a Lagrangian
  flavour and introduces integrals over paths in the configuration
  space~\cite{Feynman/Hibbs65a}.  An extension to integrals over
  phase-space paths has been done in \cite[Appendix~B]{Feynman51a}
  (see also \cite{Tobocman56a,Davies63a,Garrod66a}).}  we learn that
the value of the action is relevant beyond its stationary points all
the more than we leave the (semi-)classical domain and reach a regime
where the typical value of the action of the system is of
order~$\hbar$.  Therefore we will retain the invariance of the form of
the action as a fundamental expression of the invariance of a model:
\begin{equation}
    \T{S}_{\smash{\T{B}}}[\T{p}(\cdot),\T{q}(\cdot),\T{t}(\cdot)]=S_B[\T{p}(\cdot),\T{q}(\cdot),\T{t}(\cdot)]\;.   
\end{equation}
This means the invariance of the  boundary function up to a function of time only
\begin{equation}\label{eq:invarianceB}
      \T{B}(\T{p},\T{q},\T{t})=B(\T{p},\T{q},\T{t})+b(\T{t})
\end{equation}
and the invariance of the Hamiltonian function up to~$\dot b$
\begin{equation}\label{eq:invarianceH}
      \T{H}(\T{p},\T{q},\T{t})=H(\T{p},\T{q},\T{t})+\frac{\dmat b^{\phantom{\scriptscriptstyle\mathrm{\scriptscriptstyle T}}}}{\dmat\T{t}}(\T{t})\;
\end{equation}
that both assure the invariance of the boundary
conditions~\eqref{eq:condlim_B}.  When, on the one hand, we
put~\eqref{eq:invarianceH} into~\eqref{eq:Htransforme} and, on the
other hand, when we put~\eqref{eq:invarianceB}
into~\eqref{eq:Btransforme}, the invariance of the model under
the canonical transformation~$\mathrm{T}$ is equivalent to
\begin{equation}\label{eq:invarianceHbis}
         H(\T{p},\T{q},\T{t})\,\frac{\partial\T{t}}{\partial\Tphantom{t}}(\T{p},q,t)
         =H(p,q,t)+\frac{\partial F}{\partial t}(\T{p},q,t)\;
\end{equation}
for the Hamiltonian and
\begin{equation}\label{eq:invarianceBbis}
      B(\T{p},\T{q},\T{t})=B(p,q,t)-\T{p}\T{q}+F(\T{p},q,t)
\end{equation}
for the boundary function, once we have absorbed the irrelevant term~$b$ in an alternative
definition of~$F$.

\subsection{Conservation of the generators}

From the Hamilton's equations, the classical evolution
 of any function~$O(p,q,t)$ is given by
\begin{equation}\label{eq:dOdt}
        \frac{\dmat O}{\dmat t}=\Poisson{H}{O}+\frac{\partial O}{\partial t}\;.\\
\end{equation}
where the Poisson bracket between two phase-space functions is defined by
\begin{equation}\label{def:crochetPoisson}
  \Poisson{O_{1}}{O_{2}}\ \DEF\ 
 \partial_{p}O_{1}\partial_{q}O_{2}-\partial_{p}O_{2}\partial_{q}O_{1}
\end{equation}
(recall that the
summation/integral on the degrees of freedom is left implicit). 

Consider a continuous set of canonical transformations 
parametrised by a set of essential real
parameters~$\epsilon=(\epsilon^a)_a$ where~$\epsilon=0$ corresponds to
the identity.  The generators~$G=(G_a)_a$ of this transformation are,
by definition, given by the terms of first order in~$\epsilon$ in the
Taylor expansion of the generating function~$F(\T{p},q,t;\epsilon)$
\begin{equation}\label{eq:dldeF}
  F(\T{p},q,t;\epsilon)=\T{p}q+\epsilon\, G(\T{p},q,t)+\Omat(\epsilon^2)
\end{equation}
(in addition to the implicit summation/integral on the degrees of
freedom~$\alpha$, there is also an implicit sum on the labels~$a$ of the
essential parameters of the Lie group, those being continuous for a
gauge symmetry). We shall consider the general canonical transformations
where~$\T{t}$ is
a function of~$(\T{p},q,t)$ whose infinitesimal form is
\begin{equation}
  \T{t}(\T{p},q,t)=t+\epsilon\tau(\T{p},q,t)+\Omat(\epsilon^2). 
\end{equation}
Now with~$\T{H}(p,q,t)=H(p,q,t)$, using the form~\eqref{eq:dldeF} in
equations~\eqref{eq:Fpq} one obtains the canonical transformation
explicitly to first order
\begin{subequations}\label{subeq:Tinfinitesimal_pqtau}
\begin{eqnarray}
        \T{p}&=&p-\epsilon \partial_{q} G(p,q,t)+\epsilon H(p,q,t)\frac{\partial\tau}{\partial q}_{\;\rule[0ex]{.1ex}{3.5ex}\;p,q,t}+\Omat(\epsilon^2)\;;\\
        \T{q}&=&q+\epsilon \partial_{p} G(p,q,t)-\epsilon H(p,q,t)\frac{\partial\tau}{\partial p}_{\;\rule[0ex]{.1ex}{3.5ex}\;p,q,t}+\Omat(\epsilon^2)\;.
\end{eqnarray}
\end{subequations}
Reporting~\eqref{eq:dldeF} and~\eqref{subeq:Tinfinitesimal_pqtau}
in~\eqref{eq:invarianceHbis}, the identification of the first order
terms in~$\epsilon$ leads, with help of~\eqref{eq:dOdt},
to
\begin{equation}\label{eq:GHtauconstant}
        \frac{\dmat}{\dmat t}\Big(G(p,q,t)-\tau(p,q,t) H(p,q,t)\Big)=0.
\end{equation}
Similarly, from~\eqref{eq:invarianceBbis}, we get
\begin{equation}\label{eq:BGtauconstant}
 \tau \frac{\dmat B}{\dmat t}+\Poisson{G-\tau H}{B}+p(\partial_pG-H\partial_p\tau)-G=0
\end{equation}
where the arguments of all the functions that appear are~$(p,q,t)$.

As a special case, first consider the invariance with respect to time
translations~$\T{p}=p$, $\T{q}=q$, $\T{t}=t+\epsilon$ for any
real~$\epsilon$, then with~$F(\T{p},q,t)=\T{p}q$ corresponding to the
identity, the relations~\eqref{eq:invarianceHbis}
and~\eqref{eq:invarianceBbis} read
respectively~$H(p,q,t+\epsilon)=H(p,q,t)$ and~$B(p,q,t+\epsilon)=B(p,q,t)$
that is~$\partial_t H=0$ and~$\partial_t B=0$. The
identity~\eqref{eq:dOdt} considered for~$O=H$ and~$O=B$ leads
respectively to
\begin{equation}\label{eq:dHdtzero}
  \frac{\dmat H}{\dmat t}=0
\end{equation}
and
\begin{equation}
   \frac{\dmat B}{\dmat t}=\Poisson{H}{B}
\end{equation}
which of course are also obtained from~\eqref{eq:GHtauconstant} 
and~\eqref{eq:BGtauconstant} with~$G\equiv0$ and~$\tau\equiv1$.
Now consider a continuous set of canonical transformations such
that~$\T{t}=t$, then from~\eqref{eq:GHtauconstant} with~$\tau\equiv0$
we get
\begin{equation}\label{eq:Gconstant}
        \frac{\dmat G}{\dmat t}=0.
\end{equation}
Not only the conservation law follows straightforwardly
from~\eqref{eq:invarianceHbis} but the constant of motion are
precisely the generators of the continuous canonical
transformations~\cite{Anderson+72a}. As remarked in \cite[\S\;7
  (iv)]{Sarlet/Cantrijn81a}, Noether's result, recalled in
footnote~\ref{fn:2ndtheorem}, concerning the invariance of~$G$ under the
canonical transformations is automatic from a Hamiltonian point of view since
it follows immediately from~$\{G,G\}=0$.

Similarly,
from~\eqref{eq:BGtauconstant} with~$\tau\equiv0$ we get a relation
\begin{equation}
 \Poisson{G}{B}=G-p\,\partial_{p} G
\end{equation}
that must be fulfilled by~$B$ to have the invariance of the boundary conditions.

\section{Noether's original formulation}\label{sec:Noetherformulation}

\subsection{General variational principle}

The above result is actually completely embedded 
in Noether's original formulation except 
the discussion on the boundary conditions. Indeed, 
being more Lagrangian in flavour, \cite{Noether18a}
works systematically with a variational principle where the variations
of all the dynamical variables~$u$ vanish (as well as the derivatives of~$\delta u$ if necessary, see below).
To illustrate this 
let us first follow Noether's steps and paraphrase her analysis.
The variational principle applies to any functional whose general form is
\begin{equation}\label{def:Iu}
        S[u(\cdot)]\ \DEF\ \int_{\mathcal{D}} 
    f\big(x, u(x), \partial_x u_{|x}, \partial^2_{xx} u_{|x}, \partial^3_{xxx} u_{|x}, \dots \big)\,\dmat^d x
\end{equation}
where the functions~$u(x)=\big(u_1(x), \dots, u_N(x)\big)=\big(u_n(x)\big)_n$ (the dependent variables in Noether's terminology)
are defined on a~$d$-dimensional domain~$\mathcal{D}$ in~$\RR^d$ where some coordinates 
(the independent variables)~$x=(x^0, \dots, x^{d-1})=(x^\mu)_\mu$ are used. 
Physically, one may think the $u$'s to be various fields defined 
on some domain~$\mathcal{D}$ of space-time and~$x$ to be a particular choice of space-time coordinates.
The function~$f$ depends on~$x$, on~$u(x)$
and on their higher derivatives in~$x$ (the dots in its argument refer to 
derivatives of~$u$ of order four or more).

An infinitesimal variation~$u(x)+\delta  u(x)$ implies the first-order variation
\begin{equation}\label{eq:deltaIvariationnel}
        \delta  S\ \DEF\  S[u(\cdot)+\delta  u(\cdot)]-S[u(\cdot)]
                 = \int_{\mathcal{D}} \delta f\;\dmat^d x
\end{equation}
where~$\delta f$, with the help of integration by parts, takes the form
\begin{equation}\label{eq:deltaf}
     \delta f=\sum_{n=1}^N\  E^{n} \delta  u_n+\sum_{\mu=0}^{d-1}\  
                   \dmat_{\mu} \delta X^{\mu}=
              E\cdot \delta u
              +\dmat_x \cdot \delta X   
\end{equation}
where~$E$ stands for the~$N$-dimensional vector whose components are
\begin{equation}\label{eq:Un}
      E^n=\frac{\partial f }{\partial u_n}-\dmat_{\mu}\left( 
      \frac{\partial f }{\partial (\partial_{ \mu} u_n)}\right)
    + \dmat^2_{\mu\nu} \left(
        \frac{\partial f }{\partial (\partial^2_{ \mu\nu} u_n)}
                           \right)
    - \dmat^3_{\mu\nu\rho} \left(
        \frac{\partial f }{\partial (\partial^3_{ \mu\nu\rho} u_n)}
                           \right)
    +\cdots
\end{equation}
(from now on we will work with an implicit summation over the repeated space-time indices or field indices and the same notation ``$\,\cdot\,$'' will be indifferently used for
a --- possibly Minkowskian --- scalar product between $d$-dimensional 
space-time vectors or between~$N$-dimensional fields)
and~$\delta X$ a
$d$-dimensional  infinitesimal vector in first order in~$\delta  u$ and its derivatives
which appears through a divergence:
\begin{equation}\label{eq:deltaX}
\begin{split}
     \delta X^{\mu}= 
\left[\frac{\partial f }{\partial (\partial_{ \mu} u_n)}
   -\dmat_{\nu}\left(\frac{\partial f }{\partial (\partial^2_{ \mu\nu} u_n)}
    \right)
+\dmat^2_{\nu\rho}\left(\frac{\partial f }{\partial (\partial^3_{ \mu\nu\rho} u_n)}
    \right)-\cdots
\right]&\delta  u_n \\
+\left[ \frac{\partial f }{\partial (\partial^2_{ \mu\nu} u_n)}
  \hspace{.3cm}-\hspace{.2cm}\dmat_{\rho}\left(\frac{\partial f }{\partial (\partial^3_{ \mu\nu\rho} u_n)}
    \right)+\cdots
\right]&\partial_{ \nu}(\delta  u_n)
   \\
+\left[ \frac{\partial f }{\partial (\partial^3_{ \mu\nu\rho} u_n)}\hspace{.3cm}-\cdots\right]&\partial^2_{ \nu\rho}(\delta  u_n)
\\+&\cdots\;.
\end{split}
\end{equation}
The notation~$\dmat_{\mu}$ distinguishes the total derivative from the partial derivative~$\partial_{ \mu}$: 
\begin{equation}\label{eq:dmat_mu}
      \dmat_{\mu} =\partial_{ \mu}+ \partial_{ \mu} u_n
\frac{\partial  }{\partial  u_n}+\partial^2_{ \mu\nu}u_n
\frac{\partial  }{\partial (\partial_{ \nu} u_n)}+\cdots\;.
\end{equation}
The stationarity conditions of~$S$ when computed for the functions~$u_{\mathrm{cl}}$ imply the 
Euler-Lagrange equations
\begin{equation}\label{eq:Uclzero}
        E_{|u_{\mathrm{cl}}}=0\;.
\end{equation}
Then, remains
\begin{equation}
        \delta S[u_{\mathrm{cl}}(\cdot)]
        =\int_{\mathcal{D}}\dmat_x \cdot \delta X_{|u_{\mathrm{cl}}} \,\dmat^d x
        =\int_{\partial \mathcal{D}}\delta X_{|u_{\mathrm{cl}}}\cdot\dmat^{d-1} \sigma
\end{equation}
(Stokes' theorem leads to  the second integral which represents the outgoing flux
of the vector~$\delta X$ through the boundary~$\partial \mathcal{D}$
whose surface element is denoted by~$\dmat\sigma$)
and~$S$ will be indeed stationary if we restrict the variations~$\delta u$ on the boundaries
such that  the last integral vanishes\footnote{\label{fn:transversality}Working with~$\delta X_{|u_{\mathrm{cl}}}$
orthogonal to~$\dmat^{d-1} \sigma$ is sufficient and generalises the transversality condition
discussed in \cite[\S\S\;IV.5.2 and IV.12.9]{Courant/Hilbert53a}.
 A radical way of
  getting rid of the discussion on boundary conditions is also to work with a model
  where~$\mathcal{D}$ has no boundaries; however this is more a matter of convenience for theoreticians than a relevant feature
based on physical grounds. } (and Noether assumes that all the variations $\delta  u_n$, $\partial_{ \nu}(\delta  u_n)$, $\partial^2_{ \nu,\rho}(\delta  u_n)$ \dots
appearing in the right-hand side of~\eqref{eq:deltaX} vanish on~$\partial \mathcal{D}$).

Adding the divergence of a 
$d$-vector~$B\big(x, u(x), \partial_x u_{|x}, \partial^2_{xx} u_{|x}, \partial^3_{xxx} u_{|x}, \dots \big)$ to the integrand,
 \begin{equation}\label{eq:fBf0}
  f_B=f_0+\dmat_\mu B^\mu
\end{equation}
does not affect the expressions of the Euler-Lagrange vector~$E$
\begin{equation}
  E_B=E_0
\end{equation}
but adds to~$S$  a boundary term
\begin{equation}
  S_B[u(\cdot)]=S_0[u(\cdot)]+\int_{\partial\mathcal{D}}B\cdot\dmat^{d-1} \sigma 
\end{equation}
from which we have
\begin{equation}\label{eq:XBX0}
  \delta X_B=\delta X_0+\delta B
\end{equation}
where the infinitesimal variation~$\delta B$ comes exclusively from those of the fields~$u$ and their derivatives;
 more explicitly,
\begin{equation}
   \delta X_B^{\mu}= \delta X_0^{\mu}
+\frac{\partial B^\mu }{\partial u_n}\delta u_n
+\frac{\partial B^\mu }{\partial (\partial_\nu u_n)}\partial_\nu\delta u_n
+\frac{\partial B^\mu }{\partial (\partial^2_{\nu\rho} u_n)}\partial^2_{\nu\rho}\delta u_n
+\cdots
\end{equation}
where the ``$\cdots$'' stand for derivatives of~$B$ with respect to higher derivatives of~$u$.

\subsection{Invariance with respect to infinitesimal transformations and Noether currents}

The most general
 transformation~$\mathrm{T}$ comes with both a change of coordinates~$x\mapsto\T{x}$ and 
a change of functions~$u\mapsto\T{u}$. By definition the transformed action is given by
\begin{equation}
        \T{S}[\T{u}(\cdot)]=\int_{\Tsmall{\mathcal{D}}}
\T{f\,}\big(\T{x}, \T{u}(\T{x}), \partial_{\Tsmall{x}} \T{u}_{|\Tsmall{x}}, 
\partial^2_{\Tsmall{x}\Tsmall{x}} \T{u}_{|\Tsmall{x}}, \dots \big)\,\dmat^d  \T{x}
\end{equation}
with~$\T{S}[\T{u}(\cdot)]=S[u(\cdot)]$  for any~$u$ and for any domain~$\mathcal{D}$.
After the change of variables~$\T{x}\mapsto x$ that pulls back~$\T{\mathcal{D}}$ to~$\mathcal{D}$, we get 
\begin{multline}
    \T{f\,}\big(\T{x}, \T{u}(\T{x}), \partial_{\Tsmall{x}} \T{u}_{|\Tsmall{x}}, 
\partial^2_{\Tsmall{x}\Tsmall{x}} \T{u}_{|\Tsmall{x}}, \dots \big)
\left|\det\left(\frac{\partial \T{x}}{\partial  \Tphantom{x}}\right)\right|
\\=f\big(x, u(x), \partial_x u_{|x}, \partial^2_{xx} u_{|x}, \dots \big)
\end{multline}
which provides a \textit{definition} of~$\T{f\,}$. We have an \textit{invariance} when
the same computation rules are used to evaluate~$S$ and~$\T{S}$ that is~$ \T{f\,}=f$. Then we have
\begin{multline}\label{eq:Df}
     f\big(\T{x}, \T{u}(\T{x}), \partial_{\Tsmall{x}} \T{u}_{|\Tsmall{x}}, 
\partial^2_{\Tsmall{x}\Tsmall{x}} \T{u}_{|\Tsmall{x}}, \dots \big)
\left|\det\left(\frac{\partial \T{x}}{\partial \Tphantom{x}}\right)\right|
\\ -f\big(x, u(x), \partial_x u_{|x}, \partial^2_{xx} u_{|x}, \dots \big)
  =0\;.  
\end{multline}

The Noether conservation theorem  comes
straightforwardly from the computation of the left-hand side
of~\eqref{eq:Df} when the transformation~$\mathrm{T}$ is
infinitesimal\footnote{In Noether's spirit the transformation of all
  the dependent and independent variables can be as general as
  possible and therefore she first considers the case where~$\delta x$
  is a function of both~$x$ and~$u$; her two theorems indeed apply in
  this very general situation. Physically this corresponds to a
  transformation where the variations of the space-time
  coordinates~$\delta x$ depend not only on~$x$, as this is the case
  in General Relativity where all the diffeomorphisms of space-time
  are considered, but also on the fields~$u$. I do not know any
  relevant model in physics where this possibility has been
  exploited. In the following we will restrict~$\delta x$ to depend
  on~$x$ only, this simplification is eventually done by Noether from
  \S\;5 in \cite{Noether18a}. }:
\begin{subequations}\label{def:deltaxdeltau}
\begin{eqnarray}
\T{x}&=&x+\delta x\;;\\
\T{u}(x)&=&u(x)+\delta  u(x)\;.\label{def:deltaxdeltau_b}
\end{eqnarray}
\end{subequations} 
To first order in~$\delta x$ and~$\delta u$,~\eqref{eq:Df}
reads
 \begin{multline} \label{eq:Dfbis}
f \,\partial_x \cdot\delta x+(\partial_x f) \cdot\delta x
+\frac{\partial f }{\partial u_n} \Dsuiv u_n
+\frac{\partial f }{\partial(\partial_{ \mu} u_n)} \Dsuiv (\partial_{ \mu} u_n)
+\frac{\partial f }{\partial(\partial^2_{ \mu\nu} u_n)} \Dsuiv (\partial^2_{ \mu\nu} u_n)
+\cdots\\ \cdots +\Omat(\delta^2)=0\;,
\end{multline}
where~$\Omat(\delta^2)$ denotes terms of order at least equal to two. The first term of the 
left-hand side comes from the Jacobian
\begin{equation}
        \left|\det\left(\frac{\partial \T{x}}{\partial \Tphantom{x}}\right)\right|=
         1+\partial_x \cdot\delta x+\Omat(\delta^2)\;.
\end{equation}
The infinitesimal quantity~$\delta  u$ denotes the variation of the field~$u$
while staying at the same point~$x$ and $\Dsuiv u$ stands for the infinitesimal
variation ``following the transformation''\footnote{Borrowing the usual notation of fluid dynamics, this variation 
corresponds to the derivative following the motion often known as the convective/particle/material/Lagrangian derivative.}
\begin{equation}\label{def:Du}
        \Dsuiv u(x)\DEF \T{u}(\T{x})-u(x)=
        \delta  u(\T{x})+u(\T{x})-u(x)=
        \delta  u(x)+(\partial_x u)\cdot \delta  x+\Omat(\delta^2)\;.
\end{equation}
The chain rule for a composite function reads
\begin{equation}\label{eq:duTdxT}
        \partial_{\Tsmall{x}}\T{u}(\T{x})
        =\partial_{\Tsmall{x}}x\;\partial_x\big(\T{u}(\T{x})\big)  
       =\partial_{\Tsmall{x}}x\;\partial_x\big(u(x)+ \Dsuiv u(x)\big)
\end{equation}
where the~$d\times d$ Jacobian matrix of the transformation is
\begin{equation}\label{eq:jacobien_dxdxT}
    \partial_x{\T{x}}=\left(\partial_{\Tsmall{x}}x\right)^{-1}=
1+\partial_x\delta x+\Omat(\delta^2)\;.    
\end{equation}
By putting~\eqref{def:Du} and~\eqref{eq:jacobien_dxdxT} in~\eqref{eq:duTdxT},
we obtain\footnote{If one prefers a notation where the indices are made explicit, the
equations~\eqref{def:Ddudx} and~\eqref{def:Dd2udxdx} can be respectively re-written as
$\Dsuiv (\partial_{ \nu} u_n)=\partial_{ \nu}(\delta u_n)
   +(\partial^2_{\nu\mu}u_n)\delta x^\mu
         +\Omat(\delta^2)$ 
and
$\Dsuiv (\partial^2_{ \mu\nu} u_n)=
\partial^2_{ \mu\nu}(\delta u_n)
+(\partial^3_{\mu\nu\rho}u_n)\delta x^\rho
         +\Omat(\delta^2) \;.
$
}
\begin{equation}\label{def:Ddudx}
        \Dsuiv (\partial_x u)\DEF \partial_{\Tsmall{x}} \T{u}_{|\Tsmall{x}}-
        \partial_{x} u^{}_{|x}
=\partial_x(\delta u)+\partial_x(\partial_xu)\cdot\delta x
         +\Omat(\delta^2)   \;.
\end{equation}
In the same way, 
\begin{equation}\label{def:Dd2udxdx}
        \Dsuiv (\partial^2_{xx} u)\DEF 
        \partial^2_{\Tsmall{x}\Tsmall{x}} \T{u}_{|\Tsmall{x}}-
        \partial^2_{xx} u^{}_{|x}
         =\partial^2_{xx}(\delta u)+\partial_x(\partial^2_{xx}u)\cdot\delta x
         +\Omat(\delta^2) 
\end{equation}
and so on for the derivatives of~$u$ of higher orders.
By reporting~$\Dsuiv(\cdots)$ in~\eqref{eq:Dfbis} we get 
\begin{equation}\label{eq:Dfter}\begin{split}
        &f \,\partial_x \cdot\delta x+(\partial_x f) \cdot\delta x
+\frac{\partial f }{\partial u_n}\,(\partial_x u_n)\cdot \delta  x
\\ &+\frac{\partial f }{\partial(\partial_{ \mu} u_n)}\, \partial_x(\partial_{ \mu}   u_n)\cdot \delta  x
+\frac{\partial f }{\partial(\partial^2_{ \mu\nu} u_n)}\, \partial_x(\partial^2_{ \mu\nu} u_n)\cdot \delta  x
\\
&+\frac{\partial f }{\partial u_n}\, \delta  u_n
+\frac{\partial f }{\partial(\partial_{ \mu} u_n)}\, \partial_{ \mu} \delta  u_n
+\frac{\partial f }{\partial(\partial^2_{\mu\nu} u_n)}\, \partial^2_{ \mu\nu} \delta u_n
+\cdots+\Omat(\delta^2)=0\;.
\end{split}\end{equation} 
The first two lines provide the divergence~$\dmat_x \cdot \big(f\big(x, u(x), \partial_x u_{|x}, \partial^2_{xx} u_{|x}, \dots \big)\delta x\big)$ and at the last line we recognise the variation~$\delta  f$ given by~\eqref{eq:deltaf}.
Then
\begin{equation}\label{eq:Udeltau_ddeltaJ}
        E\cdot \delta u
              +\dmat_x \cdot (\delta X+f\delta x)=0\;.
\end{equation}
With the help of~\eqref{eq:Uclzero}, we deduce Noether's conservation law for the infinitesimal current:
\textit{
If the functional~\eqref{def:Iu} is invariant under a continuous family of transformations 
having, in the neighbourhood of the identity the form~~\eqref{def:deltaxdeltau}, then for any solution~$u_{\mathrm{cl}}$ such that~$S$
is stationary, the  (infinitesimal) Noether current
 \begin{equation}
         \delta J\ \DEF\ \delta X+f\delta x
\end{equation}
with~$\delta X$ given by~\eqref{eq:deltaX} is conserved; that is
\begin{equation}\label{eq:conservationdeltaJ}
    \dmat_x \cdot  \delta J_{|u_{\mathrm{cl}}}=\dmat_{\mu} \delta J^\mu_{|u_{\mathrm{cl}}}=0    \;.
\end{equation} 
}
\noindent More explicitly we have
\begin{subequations}
\begin{align}
     &\delta J^{\mu}  
=f\delta  x^\mu+\frac{\partial f }{\partial (\partial_{ \mu} u_n)}\delta  u_n
   -\dmat_{\nu}\left(\frac{\partial f }{\partial (\partial^2_{ \mu\nu} u_n)}
    \right)\delta  u_n + \frac{\partial f }{\partial (\partial^2_{ \mu\nu} u_n)}\partial_{ \nu}(\delta  u_n)
    +\cdots \label{eq:deltaJdeltau}\\
&= \left[f\updelta^{\mu}_{\nu}-\frac{\partial f }{\partial (\partial_{ \mu} u_n)}\partial_{ \nu} u_n
     +\dmat_{\rho}\left(\frac{\partial f }{\partial (\partial^2_{ \mu\rho} u_n)}
    \right)\partial_{ \nu} u_n
     - \frac{\partial f }{\partial (\partial^2_{ \mu\rho} u_n)}
   \partial^2_{ \nu\rho} u_n +\cdots
  \right] \delta x^\nu
 \nonumber\\ 
& +\frac{\partial f }{\partial (\partial_{ \mu} u_n)}\Dsuiv u_n
  -\dmat_{\nu}\left(\frac{\partial f }{\partial (\partial^2_{ \mu\nu} u_n)}
 \right)\Dsuiv u_n + \frac{\partial f }{\partial (\partial^2_{ \mu\nu} u_n)}\Dsuiv(\partial_{ \nu} u_n)
    +\cdots  
\end{align}
\end{subequations}
where the Kronecker symbol~$\updelta$ is used and 
``$\cdots$''  stands for terms involving the derivatives of~$f$ with respect to third order or higher derivatives 
of~$u$.  Since the invariance of the variational
problem depends on the choice of the boundary function, so will the Noether current
as we can see from~\eqref{eq:fBf0} and~\eqref{eq:XBX0}:
\begin{equation}\label{eq:JBJ0}
   \delta J_B=\delta J_0+(\dmat_x \cdot  B) \delta x+\delta B\;.
\end{equation}
In fact, the Noether infinitesimal currents~$\delta J$ are defined up to a divergence-free current since adding such a term does not
 affect~\eqref{eq:conservationdeltaJ}. For instance 
\begin{equation}
  \delta J'^\mu=\delta J^\mu+\dmat_\nu\left[\left(
\frac{\partial B^\mu }{\partial (\partial_\nu u_n)}
-\frac{\partial B^\nu }{\partial (\partial_\mu u_n)}
\right)\delta u_n\right]
\end{equation}
 would  also be an acceptable Noether current associated with the symmetry under the scope.

\subsection{Aside remarks about the two Noether theorems}\label{subsec:2theorems}

The
result established in the previous section
 is neither the first Noether theorem nor the second one 
but encapsulates both of
them; the conservation of the \textit{infinitesimal} current~$\delta J$
occurs for any global or local symmetry. 
Noether's first theorem follows 
from the computation of~$\delta X$ for a global symmetry i.e. when
the number of the essential parameters~$\epsilon=(\epsilon^a)_a$
 of the Lie group of transformations is finite. In that case
\begin{equation}\label{deltaJpropepsilon}
  \delta J=\mathscr{J}\epsilon+\Omat(\epsilon^2)
\end{equation}
or in terms of coordinates
\begin{equation}\label{deltaJpropepsiloncoord}
  \delta J^\mu=\mathscr{J}_a^\mu\epsilon^a+\Omat(\epsilon^2)
\end{equation}
and the first Noether theorem states the
 conservation of the \textit{non infinitesimal}~$\mathscr{J}$
\begin{equation}
  \dmat_x\cdot\mathscr{J}_a=\partial_\mu\mathscr{J}^\mu_a=0
\end{equation}
obtained immediately from the infinitesimal conservation 
law~\eqref{eq:conservationdeltaJ} since~$\epsilon$ is arbitrary and $x$-independent.
   
Noether's second theorem (see 
footnote~\ref{fn:2ndtheorem}) 
follows 
from the computation of~$\delta X$
for a local symmetry i.e. when the essential parameters are functions~$\epsilon(x)$
and, in that case, the proportionality relation~\eqref{deltaJpropepsilon} 
does not hold anymore ; the right-hand side now includes the derivatives of~$\epsilon$:  
\begin{equation}
        \delta J^{\mu}=\mathscr{J}^{\mu}\epsilon
       + {\mathscr{F}}^{\mu\nu}\partial_{\nu}\epsilon
        + {\mathscr{K}}^{\mu\nu\rho}\partial^2_{\nu\rho}\epsilon       
       +\cdots+\Omat(\epsilon^2)\;.
\end{equation}
By expanding the variation of the fields according to
\begin{equation}
        \delta u=\frac{\partial u}{\partial \epsilon}\epsilon
         +\frac{\partial u}{\partial(\partial_{\mu}\epsilon)}\,\partial_{\mu}\epsilon
         +\frac{\partial^2u}{\partial(\partial^2_{\mu\nu}\epsilon)}\,\partial^2_{\mu\nu}\epsilon+\cdots
+\Omat(\epsilon^2)\;,
\end{equation}
then~\eqref{eq:Udeltau_ddeltaJ} reads
\begin{multline}\label{eq:expansionenderiveesepsilon}
        \left[E\cdot\frac{\partial u}{\partial \epsilon}+\dmat_\mu\mathscr{J}^{\mu}\right]\epsilon+
        \left[E\cdot\frac{\partial u}{\partial(\partial_{ \mu}\epsilon)}
         +\mathscr{J}^{\mu}+\dmat_\nu {\mathscr{F}}^{\nu\mu}
        \right]\partial_{ \mu}\epsilon\\
        +\left[ E\cdot\frac{\partial^2 u}{\partial(\partial^2_{ \mu\nu}\epsilon)}
        +\frac{1}{2}({\mathscr{F}}^{\nu\mu}+{\mathscr{F}}^{\mu\nu})
        +\dmat_\rho {\mathscr{K}}^{\rho\nu\mu}
        \right]\partial^2_{ \mu\nu}\epsilon
+\cdots=0\;.
\end{multline}
Since the functions~$\epsilon$ are arbitrary, all the brackets vanish separately.
When evaluated on the stationary solutions~$u_{\mathrm{cl}}$, we get
\begin{equation}\label{eq:identitecourantsjauge}
        \dmat_\mu\mathscr{J}^{\mu}=0\;;\quad
        \dmat_\mu {\mathscr{F}}^{\mu\nu}=-\mathscr{J}^{\nu}\;,\quad
\dmat_\rho {\mathscr{K}}^{\rho\nu\mu}=-\frac{1}{2}({\mathscr{F}}^{\nu\mu}+{\mathscr{F}}^{\mu\nu})\;,
\quad\text{etc.}
\end{equation}
For a constant~$\epsilon$ we recover the first theorem from the first equality. 
The second theorem stipulates
that to each~$a$ there is one identity connecting the~$E$'s:
\begin{equation}\label{eq:Bianchi}
  E\cdot\frac{\partial u}{\partial \epsilon}-\dmat_\mu\left(E\cdot\frac{\partial u}{\partial(\partial_{ \mu}\epsilon)}\right)+\dmat_\mu\dmat_\nu\left(E\cdot\frac{\partial^2 u}{\partial(\partial^2_{ \mu\nu}\epsilon)}\right)+\cdots=0\;. 
\end{equation}
 Those can be obtained from the 
vanishing brackets of~\eqref{eq:expansionenderiveesepsilon} or directly from the following
re-writing of~\eqref{eq:Udeltau_ddeltaJ}: 
\begin{equation}\label{eq:Udeltau_ddeltaJbis}
\begin{split}
        &\left[E\cdot\frac{\partial u}{\partial \epsilon}\right.\left.-\dmat_\mu\left(E\cdot\frac{\partial u}{\partial(\partial_{ \mu}\epsilon)}\right)+\dmat_\mu\dmat_\nu\left(E\cdot\frac{\partial^2 u}{\partial(\partial^2_{ \mu\nu}\epsilon)}\right)
        +\cdots\right]\epsilon\\
        +\dmat_\mu  &\left[\delta J^\mu
        +E\cdot\frac{\partial u}{\partial(\partial_{ \mu}\epsilon)}\,\epsilon
        -\dmat_\nu\left( E\cdot\frac{\partial^2 u}{\partial(\partial^2_{ \mu\nu}\epsilon)}\right)\epsilon
        +E\cdot\frac{\partial^2 u}{\partial(\partial^2_{ \mu\nu}\epsilon)}\,\partial_{ \nu}\epsilon
        +\cdots\right]=0\;. \raisetag{3\baselineskip}
\end{split}
\end{equation}
By an integration on any arbitrary volume and choosing~$\epsilon$ and its derivatives
vanishing on its boundary, on can get rid of
the integral of the second term of the left-hand-side. Since~$\epsilon$ can be chosen
otherwise arbitrarily within this volume, the first bracket vanishes which is exactly
the Noether identity~\eqref{eq:Bianchi}\footnote{As a consequence, the cancellation 
of the first bracket in~\eqref{eq:expansionenderiveesepsilon} allows to write the first term of~\eqref{eq:Bianchi} as a total derivative
and this leads to the conservation 
of a current 
\begin{equation}
  \dmat_\mu  \left[\mathscr{J}^{\mu} 
        +E\cdot\frac{\partial u}{\partial(\partial_{ \mu}\epsilon)}
        -\dmat_\nu\left( E\cdot\frac{\partial^2 u}{\partial(\partial^2_{ \mu\nu}\epsilon)}\right)
        +\cdots\right]=0
\end{equation}
which is qualified as 
a ``strong'' \cite[\S\;6 and its references]{Barbashov/Nesterenko83a} because this constraint
holds even if the Euler-Lagrange equations are not satisfied (a primary constraint in Dirac's terminology \cite{Dirac64a}).
}.  If one had to speak of just one theorem connecting symmetries and conservation laws, 
one could choose the cancellation of all the brackets of~\eqref{eq:expansionenderiveesepsilon} 
from which Noether's theorems~I and~II are particular cases.

Eventually, let us mention that both Noether's theorems include also a reciprocal statement:
the invariance in the neighbourhood of~$\epsilon=0$ implies an invariance 
for any finite~$\epsilon$ and this comes from the properties of the
underlying Lie structure of the transformation group and its internal composition law
that allow to 
naturally map any neighbourhood of~$\epsilon=0$ to a neighbourhood of any other element
of the group.

\section{Applications}

\subsection{Finite number of degrees of freedom}\label{subsec:finitedof}

From the general 
formalism in \S\;\ref{sec:Noetherformulation} it is straightforward
to show  that the conservation law we obtained within the Hamiltonian framework
 in~\S\;\ref{subsec:clinvariance} is encapsulated in Noether's original approach.
For~$\Deg$ degrees of freedom~$q=(q_\alpha)_{\alpha\in\finiteset{1}{\Deg}}$ we have~$u=(p,q,t)$ with~$N=2\Deg+1$, 
$S$ is of course~$S_B$ given by equation~\eqref{def:actionhamiltonienneB}, 
$\mathcal{D}$ is~$[s_i,s_f]$, $x$ is identified with~$s$
($d=1$) and only the first derivatives of~$q$, $t$,
and possibly~$p$ through~$\dmat B/\dmat s$ are involved. We are considering transformations where~$s$ is unchanged:
 $\delta x=\delta s=0$, and then, $\Dsuiv=\delta$. Therefore, 
with~$f_B(p,q,t,\dmat p/\dmat s,\dmat q/\dmat s,\dmat t/\dmat s)=p\,\dmat q/\dmat s-H(p,q,t)\,\dmat t/\dmat s+\dmat B/\dmat s$, 
\begin{subequations}
\begin{eqnarray}
        \delta J_B= \delta X_B
         &=&\frac{\partial f_B }{\partial (\dmat p/\dmat s)}\delta  p
          +\frac{\partial f_B }{\partial (\dmat q/\dmat s)}\delta  q
          +\frac{\partial f_B }{\partial (\dmat t/\dmat s)}\delta  t\;;\\
         &=&\partial_{ p}B\,\delta  p +(p+\partial_{ q}B)\,\delta  q-
(H-\partial_{ t}B)\,\delta  t\;;\\
      &=&p\delta  q-H\delta  t+\delta B\;,\label{eq:deltaJ_ham}
\end{eqnarray}
\end{subequations}
which is a particular case of~\eqref{eq:JBJ0}.
Precisely because of the invariance, 
the variations coming from the infinitesimal transformation under the scope 
naturally satisfy the boundary conditions  \eqref{eq:condlim_B} used to formulate the variational principle  that now can be interpreted
as the conservation of~$\delta J_B$ between~$s_i$ and~$s_f$.
The invariance~\eqref{eq:invarianceBbis} of~$B$ reads
$\delta B=-\T{p}\T{q}+F(\T{p},q,t)=-\T{p}(\T{q}-q)+\epsilon G(\T{p},q,t)+\Omat(\epsilon^2)
=-p\delta  q+\epsilon G(p,q,t)+\Omat(\epsilon^2)$ and
\begin{equation}\label{eq:deltaJeGHdeltat}
         \delta J_B= \epsilon G(p,q,t) -H\delta  t\;.
\end{equation}
For an arbitrary pure time translation~$\delta t$ is $s$-independent
and~$\epsilon=0$, then~\eqref{eq:conservationdeltaJ}, which
reads~$\dmat \delta J_B/\dmat s=0$, just expresses the constancy
of~$H$. For a canonical transformation that does not affect the time,
the latter equation shows that its generator~$G$ is an integral of
motion. Thus, with a presentation much closer to Noether's original
spirit we actually recover the results of
section~\S\;\ref{subsec:clinvariance}.  What is remarkable is that, in
the latter case, the Noether constants are independent of~$H$ and~$B$ whereas,
a priori, the general expression of the
current~\eqref{eq:deltaJdeltau} depends on~$f$ (see also~\eqref{eq:JBJ0}): only the canonical
structure, intimately bound to the structure of the
action~\eqref{def:actionhamiltonienne}, leaves its imprint whereas the
explicit forms of the Hamiltonian and the boundary function have
 no influence on the expression of
the conserved currents (as soon as the invariance is maintained of
course). In other words, it is worth noticed that the Noether currents
keep the same expression for all the (infinite class of) actions that
are invariant under the associated transformations.

\subsection{Examples in field theory}\label{subsec:field}

The discussion of the previous paragraph still 
holds at the limit~$\Deg\to\infty$
but it is worth to adapt it to the case of  field models.
A field involves an infinite number of degrees of freedom that
we shall take continuous and preferably labelled by the
$\Dim$-dimensional space coordinates~$\alpha=\tv{x}$ rather than the
dual wave-vectors~$\tv{k}$.  The additional discrete ``internal''
quantum numbers like those that distinguish the spin components are
left implicit.  Now the Hamiltonian appears to be a functional of the
dynamical variables, namely the fields~$\{\pi,\varphi\}$ and their
spatial derivatives---restricted to order one for the sake of
simplicity whereas we have seen from the general approach that this
assumption is not mandatory---of the form
\begin{equation}\label{eq:hamdensity}
      H[\pi(t,\cdot),\varphi(t,\cdot),t]=\int_{\mathcal{V}} \mathscr{H}\big(\pi(t,\tv{x}),\varphi(t,\tv{x}),\grads\pi(t,\tv{x}),\grads\varphi(t,\tv{x}),t,\tv{x} \big)\,\dmat^\Dim\tv{x}
\end{equation}
where~$\mathcal{V}$ is a $\Dim$-dimensional spatial domain and~$\mathscr{H}$, 
the Hamiltonian density that may a priori depend explicitly on~$x=(t,\tv{x})$.
The action 
\begin{equation}\label{eq:Schamp}
  S_B[\pi(\cdot),\varphi(\cdot)]=\int_{\mathcal{V}\times[t_i,t_f]}\left(\pi\partial_{ t}\varphi-\mathscr{H}+\dmat_t\mathscr{B}\right)\dmat^\Dim\tv{x}\, \dmat t
\end{equation}
involves a boundary density~$\mathscr{B}\big(\pi(t,\tv{x}),\varphi(t,\tv{x}),\grads\pi(t,\tv{x}),\grads\varphi(t,\tv{x}),t,\tv{x} \big)$
from which the boundary function(nal) is given by~$\int_{\mathcal{V}}\mathscr{B}\dmat^\Dim\tv{x}$ 
keeping the same locality principle as we used for~$H$ 
(we assume that neither~$H$ nor~$B$ involve non-local terms like~$\varphi(\tv{x})V(\tv{x}'-\tv{x})\varphi(\tv{x})$). Whenever working in a relativistic framework,
$\mathscr{B}$ can be seen as the 0th-component of a~$(\Dim+1)$-vector $B=(B^0,B^1,\dots,B^\Dim)=(\mathscr{B},0,\dots,0)$
such that~$\dmat_t\mathscr{B}=\dmat_\mu B^\mu$ and the space-time integral defining~$S_B$ can be seen as an integral over
the~$d=(\Dim+1)$-dimensional domain~$\mathcal{D}$ between two appropriate Cauchy surfaces. The action~\eqref{eq:Schamp} takes the
general expression form~\eqref{def:Iu} 
with now~$N=2$ fields~$u=(u_1,u_2)=\big(\pi,\varphi\big)$ and~$f$ given by
\begin{equation}
  f_B\big(\pi,\varphi,\partial_x\pi,\partial_x\varphi,x \big)=\pi\partial_0\varphi-\mathscr{H}\big(\pi,\varphi,\grads\pi,\grads\varphi,x \big)+\dmat_\mu B^\mu\;.
\end{equation}
By cancelling the components~$E^1$ and~$E^2$ computed from~\eqref{eq:Un} we obtain the evolution equations of the classical fields 
\begin{subequations}\label{eq:Hamiltondensite}
        \begin{eqnarray}
        \partial_t \varphi&=&\phantom{-}
\frac{\partial \mathscr{H}}{\partial\pi}-
\frac{\dmat}{\dmat x^i}\left(\frac{\partial \mathscr{H}}{\partial(\partial_i\pi)}\right)\;;
\label{eq:Hamiltondensite_phidot}\\
        \partial_t \pi&=&-\frac{\partial \mathscr{H}}{\partial\varphi}+
\frac{\dmat}{\dmat x^i}\left(\frac{\partial \mathscr{H}}{\partial(\partial_i\varphi)}\right)\;.\label{eq:Hamiltondensite_pidot}
        \end{eqnarray}
\end{subequations}
The Noether infinitesimal current is given by~\eqref{eq:JBJ0} with
\begin{equation}\label{eq:deltaJmu_H}
\begin{split}
         \delta J_B^{\mu}=&(\pi\partial_{ t}\varphi-\mathscr{H})\delta  x^\mu
         +\left(\pi\updelta^\mu_0-\frac{\partial \mathscr{H} }{\partial (\partial_{ \mu} \varphi)}\right)\delta \varphi
         -\delta \pi\frac{\partial \mathscr{H} }{\partial (\partial_{ \mu} \pi)}\\
&+(\dmat_\rho B^\rho) \delta  x^\mu+\frac{\partial B^\mu}{\partial \varphi^{\;}}\delta \varphi
+\delta \pi\frac{\partial B^\mu}{\partial \pi^{\;}}
+
\frac{\partial B^\mu }{\partial (\partial_\rho \varphi)}\partial_\rho\delta\varphi
+\partial_\rho\delta\pi
\frac{\partial B^\mu }{\partial (\partial_\rho \pi)}\;.
\end{split}\raisetag{3.3\baselineskip}
\end{equation}

As an illustration, let us specify the latter general expression in the special
case of the space-time translations.  
We have~$\T{\varphi}(x)=\varphi(x-\delta x)$ and $\T{\pi}(x)=\pi(x-\delta x)$ so the infinitesimal variations
of the fields are 
\begin{equation}
        \delta \pi=-\partial\pi\cdot\delta x\;;\qquad
        \delta \varphi=-\partial\varphi\cdot\delta x
\end{equation}
and then, since we take~$\delta x$ to be independent of~$x$, 
we get (\textit{cf} equation~\eqref{deltaJpropepsiloncoord} with~$a$ being now the
space-time label and~$\epsilon=\delta x$)  
\begin{equation}
  \delta J^\mu_B=\mathscr{T}_{B|\nu}^{\mu}\delta x^\nu
\end{equation}
with the energy-momentum tensor given up to a divergence-free current\footnote{Adding a divergence-free current 
may be exploited to work with a symmetric tensor known as the Belinfante-Rosenfeld tensor
since this was first proposed by \cite{Belinfante39a,Rosenfeld40a}.}  by
\begin{eqnarray}
  \mathscr{T}_{B|\nu}^{\mu}\!\!&=&\!\mathscr{T}_{0|\nu}^{\mu}\!+\!(\dmat_\rho B^\rho)\updelta^\mu_\nu
-\frac{\partial B^\mu}{\partial \varphi^{\;}}\partial_\nu\varphi
-\partial_\nu\pi\frac{\partial B^\mu}{\partial \pi^{\;}}
-\frac{\partial B^\mu }{\partial (\partial_\rho\varphi)}\partial^2_{\rho\nu}\varphi
-\partial^2_{\rho\nu}\pi\frac{\partial B^\mu }{\partial (\partial_\rho \pi)}\;;\nonumber \\ \\
&=&\!\mathscr{T}_{0|\nu}^{\mu}\!+\!(\dmat_\rho B^\rho)\updelta^\mu_\nu-\dmat_\nu B^\mu+\partial_\nu B^\mu\;,
\end{eqnarray}
and
\begin{equation}
   \mathscr{T}_{0|\nu}^{\mu}=(\pi\partial_{ t}\varphi-\mathscr{H})\,\updelta^\mu_\nu
                   +\partial_{ \nu}\pi\frac{\partial \mathscr{H} }{\partial (\partial_{ \mu} \pi)}+\Big(\frac{\partial \mathscr{H} }{\partial (\partial_{ \mu} \varphi)}-\pi\updelta^\mu_0\Big)\partial_{ \nu}\varphi\;.      
\end{equation}
The invariance of the boundary function under translations requires~$\partial_\nu B^\mu=0$ and 
the corresponding $(\Dim+1)$-momentum contained in the volume~$\mathcal{V}$ is therefore given by
\begin{equation}
  P_{B|\nu}=\int_{\mathcal{V}} \mathscr{T}_{B|\nu}^{0}\;\dmat^\Dim\tv{x}=P_{\nu}+\Delta P_\nu
\end{equation}
where
\begin{equation}
  P_{\nu}=\int_{\mathcal{V}}\Big[(\pi\partial_{ t}\varphi-\mathscr{H})\,\updelta^0_\nu-\pi\partial_{ \nu}\varphi\Big]\;\dmat^\Dim\tv{x}\;.
\end{equation}
On can check that~$P^0=-P_{0}$ is given by~\eqref{eq:hamdensity}. The boundary function brings some
surface corrections
\begin{equation}
  \Delta P_\nu= \int_{\mathcal{V}}\Big[(\dmat_\rho B^\rho)\updelta^0_\nu-\dmat_\nu B^0
\Big]\;\dmat^\Dim\tv{x}
\end{equation}
that is
\begin{equation}
  \Delta P_0= \int_{\mathcal{V}}\dmat_i B^i \;\dmat^\Dim\tv{x}=\int_{\partial\mathcal{V}} B^i \;\dmat^{\Dim-1}\sigma_i
\end{equation}
and
\begin{equation}
  \Delta P_i=\int_{\mathcal{V}}\dmat_i B^0 \dmat^\Dim\tv{x}=\int_{\partial\mathcal{V}}B^0 \;\dmat^{\Dim-1}\sigma_i
\end{equation}
where~$\dmat^{\Dim-1}\sigma_i$ are the~$\Dim$ components of the surface element defined on~$\partial\mathcal{V}$.
In any reasonable model these corrections are expected to vanish when~$\partial\mathcal{V}$ is extended to infinity.

\subsection{Comparison with the Lagrangian approach}\label{subsec:lagrangian}

For the sake of completeness let us comment on the connection with the
Lagrangian framework of a system with~$\Deg$ degrees of
freedom. Consider now~\eqref{def:Iu} with $f$ being $L(q,\dot q,
t)+\dmat B/\dmat t$ where~$B$ is a function of~$q$, $\dot{q}$ and~$t$,
the integration variable~$x$ is just the time~$t$ ($d=1$) and the
number of dynamical variables $u=q$ is divided by two ($N=\Deg$) by
comparison with the Hamiltonian framework. The derivative~$\dmat
B/\dmat t$ depends on~$\ddot{q}$ and this must be taken into account
when computing directly~$\delta X$ from~\eqref{eq:deltaX}
\begin{subequations}
\begin{align}
        \delta X&=\frac{\partial f}{\partial\dot{q}}\delta q
                 -\frac{\dmat}{\dmat t}\left(\frac{\partial f }{\partial \ddot{q}}\right)\delta q
                 + \frac{\partial f }{\partial \ddot{q}}\frac{\dmat \delta q}{\dmat t}\;;\\
                 &=\Bigg[\frac{\partial L}{\partial\dot{q}}+\frac{\partial B}{\partial q}
                    \underbrace{+\frac{\partial^2 B}{\partial\dot{q}\partial\dot{q}}\ddot{q}
                    +\frac{\partial^2 B}{\partial\dot{q}\partial q}\dot{q}
                    +\frac{\partial^2 B}{\partial\dot{q}\partial t}
                    - \frac{\dmat}{\dmat t}\left(\frac{\partial B}{\partial \dot{q}}\right)}_{\displaystyle =0}\Bigg]\delta q
                    +\frac{\partial B}{\partial \dot{q}}\frac{\dmat\delta q}{\dmat t}.
\end{align}
\end{subequations}
Hence, since~$\delta x=\delta t$, we have rederived a particular case of~\eqref{eq:JBJ0},
\begin{equation}\label{eq:deltaJ_lag}
        \delta J=\frac{\partial L}{\partial\dot{q}}\delta q+L\delta t
                +\frac{\partial B}{\partial q}\delta q+\frac{\partial B}{\partial \dot{q}}\frac{\dmat\delta q}{\dmat t}+\frac{\dmat B}{\dmat t}\delta t\;.
\end{equation}
To reconcile~\eqref{eq:deltaJ_lag} and~\eqref{eq:deltaJ_ham}, 
one must be aware that~$\delta  q$ has a different meaning in the two equations. Indeed,
in the general expression~\eqref{eq:deltaX}~$\delta  u$ stands for
a variation of~$u$ computed at the same~$x$ 
(see~\eqref{def:deltaxdeltau_b});  within the Hamiltonian formalism, 
$\delta ^{\scriptscriptstyle(\mathrm{ham})} q$ thus denotes a variation of~$q$ 
at the same parameter~$s$ whereas within the Lagrangian formalism, $\delta ^{\scriptscriptstyle(\mathrm{lag})} q$ denotes a   
variation of~$q$ at the same time~$t$. Precisely when the transformation modifies~$t$, these two variations differs. To connect them one has to introduce the parametrisation~$s$
in the Lagrangian formalism
\begin{equation}
         \delta ^{\scriptscriptstyle(\mathrm{lag})} q\big(t(s)\big)= \T{q}\big(t(s)\big)-q\big(t(s)\big)
\end{equation}
and then
\begin{equation}
         \delta ^{\scriptscriptstyle(\mathrm{ham})} q\big(t(s)\big)=\T{q}\big(\T{t}(s)\big)-q\big(t(s)\big)=D^{\scriptscriptstyle(\mathrm{lag})}q\;,
\end{equation}
with~$\T{t}(s)=t(s)+\delta t(s)$
\footnote{Because
$\delta ^{\scriptscriptstyle(\mathrm{ham})}t
=\T{t}(s)-t(s)=\T{t}-t=\delta ^{\scriptscriptstyle(\mathrm{lag})}t$, 
we will not use two different notations for the variations of~$t$.}. 
Then,
\begin{equation}
        \delta ^{\scriptscriptstyle(\mathrm{lag})} q=\delta ^{\scriptscriptstyle(\mathrm{ham})} q-\dot{q}\delta t\;.
\end{equation}
Reporting this last expression in~\eqref{eq:deltaJ_lag}, we get
\begin{equation}
\delta J=\frac{\partial L}{\partial\dot{q}}\delta ^{\scriptscriptstyle(\mathrm{ham})}q+\left(L-\frac{\partial L}{\partial\dot{q}}\dot{q}\right)\delta t
 +\frac{\partial B}{\partial t}\delta t
 +\frac{\partial B}{\partial q}\delta ^{\scriptscriptstyle(\mathrm{ham})}q
 +\frac{\partial B}{\partial \dot{q}}
  \left(\frac{\dmat\delta ^{\scriptscriptstyle(\mathrm{ham})}q}{\dmat t}-\dot{q}\frac{\dmat\delta t }{\dmat t}\right)\;. 
\end{equation} 
Turning back to the parametrisation by~$s$, the last parenthesis is
\begin{eqnarray}
 \delta ^{\scriptscriptstyle(\mathrm{ham})}\left(\frac{\dmat q}{\dmat t}\right)
 &=&\delta ^{\scriptscriptstyle(\mathrm{ham})}\left(\frac{1}{\dmat t/\dmat s}\frac{\dmat q}{\dmat s}\right)\;;\\ 
&=&
        \frac{1}{\dmat t/\dmat s}\underbrace{\delta ^{\scriptscriptstyle(\mathrm{ham})}\left(\frac{\dmat q}{\dmat s}\right)}_{\displaystyle=\frac{\dmat\delta ^{\scriptscriptstyle(\mathrm{ham})} q}{\dmat s}}-\frac{1}{(\dmat t/\dmat s)^2}
      \frac{\dmat q}{\dmat s}\underbrace{\delta ^{\scriptscriptstyle(\mathrm{ham})}\left(\frac{\dmat t}{\dmat s}\right)}_{\displaystyle=\frac{\dmat\delta  t}{\dmat s}}\;.
\end{eqnarray}
therefore one recovers
\begin{equation}
 \delta J=\frac{\partial L}{\partial\dot{q}}\delta ^{\scriptscriptstyle(\mathrm{ham})}q+\left(L-\frac{\partial L}{\partial\dot{q}}\dot{q}\right)\delta t
 +\delta ^{\scriptscriptstyle(\mathrm{ham})}B
\end{equation}
which coincides with~\eqref{eq:deltaJ_ham} using 
\begin{equation}\label{def:lagrangien}
     L\left( q,\frac{\dmat q}{\dmat t}, t\right)
     \ \DEF\  p\,\frac{\dmat q}{\dmat t}-H(p,q,t)\;.  
\end{equation}
We could also have obtained~\eqref{eq:deltaJ_ham} by working with the
Lagrangian functional where all the functions are systematically computed with~$s$
\begin{equation}
       S_B=\int_{s_i}^{s_f}\left[L\left(q(s),\frac{1}{\dmat t/\dmat s}\frac{\dmat q}{\dmat s},t(s)\right)\frac{\dmat t}{\dmat s}+\frac{\dmat B}{\dmat s}\right]\dmat s \;,
\end{equation}
or, conversely, by eliminating all the references to~$s$ in the Hamiltonian functional
\begin{equation}
       S_B=\int_{t_i}^{t_f}\left[p\frac{\dmat q}{\dmat t}-H+\frac{\dmat B}{\dmat t}\right]\dmat t\;. 
\end{equation}

For a Lagrangian field model we have $u=\varphi$ and~\eqref{eq:deltaJdeltau}
reads
\begin{equation}\label{eq:deltaJmu_L}
        \delta J^{\mu}=\mathscr{L}\delta  x^\mu+
        \frac{\partial \mathscr{L} }{\partial (\partial_{ \mu} \varphi)}\delta \varphi\;.
\end{equation}
Using the fact that~$\mathscr{H}$ does not depend on~$\partial_t\pi$ 
nor~$\partial_t\varphi$ and with the help of
\begin{equation}\label{eq:densites_HL}
        \mathscr{L}=\pi\partial_t\varphi-\mathscr{H}
\end{equation}
and
\begin{equation}\label{eq:piL}
       \pi=\frac{\partial \mathscr{L}}{\partial(\partial_t\varphi)}\;,
\end{equation}
 the two first terms
of the right-hand-side of~\eqref{eq:deltaJmu_H} are identical to those appearing 
in~\eqref{eq:deltaJmu_L}:
\begin{equation}
         \delta J_0^{\mu}=(\pi\partial_{ t}\varphi-\mathscr{H})\delta  x^\mu
         +\left(\pi\updelta^\mu_0-\frac{\partial \mathscr{H} }{\partial (\partial_{ \mu} \varphi)}\right)\delta \varphi
         -\delta \pi\frac{\partial \mathscr{H} }{\partial (\partial_{ \mu} \pi)}\;.
\end{equation}
The two currents coincide when~$\mathscr{H}$ does not depend on~$\grads\pi$ which is
a standard case. 

\section{Quantum framework} \label{sec:quantum}

\subsection{Complex canonical formalism}

In quantum theory, any state~$\ket{\psi}$ can be represented by the 
list $z=(z_\alpha)_\alpha$ of its complex components~$z_\alpha \DEFt\braket{\phi_\alpha}{\psi}$
on a given orthonormal basis~$\big(\ket{\phi_\alpha}\big)_\alpha$ labelled by the quantum numbers~$\alpha$.
For simplicity we will work with discrete quantum numbers but this is not
a decisive hypothesis here and what follows can be adapted to relativistic as well as non-relativistic 
quantum field theory. The quantum evolution is governed
by a self-adjoint Hamiltonian\footnote{For a non-isolated system, even in the Schr\"odinger picture, the Hamiltonian may depend on time.}~$\hat{H}(t)$ according to 
\begin{equation}\label{eq:evolutionket}
  \imat\hbar\frac{\dmat}{\dmat t}\ket{\psi}=\hat{H}(t)\ket{\psi}
\end{equation} 
or equivalently
\begin{equation}\label{eq:schrocomplexe_z}
        \imat\hbar \dot{z}_\alpha(t)=
       \sum_{\alpha'}H_{\alpha,\alpha'}(t)\;z_{\alpha'}(t)
\end{equation}
with the matrix element
\begin{equation}
        H_{\alpha,\alpha'}(t)\DEF\bra{\phi_\alpha}\hat{H}(t)\ket{\phi_{\alpha'}}\,.
\end{equation}
Provided we accept to extend the classical Hamiltonian formalism to complex dynamical variables, one can see that
the quantum dynamics described above can be derived from the ``classical'' quadratic Hamiltonian
\begin{equation}\label{def:Hwz}
  \mathsf{H}(w,z,t)\DEF\frac{1}{\imat\hbar}\sum_{\alpha,\alpha'}w_\alpha H_{\alpha,\alpha'}(t) z_{\alpha'}
\end{equation}
where each couple~$(w_\alpha,z_\alpha)$ is now considered as a pair of complex canonical variables~$(p_\alpha,q_\alpha)$\footnote{And it is worth noting that the complex Poisson bracket between two quadratic functions defined
like~\eqref{def:Hwz}
from two operators  keeps the same form and is expressed
with the matrix elements of the commutator of the two operators.}.
The equation~\eqref{eq:schrocomplexe_z} corresponds to Hamilton's equations for~$q$ whereas Hamilton's equations for~$p$
are
\begin{equation}\label{eq:schrocomplexe_zbar}
        \imat\hbar \dot{w}_\alpha(t)=-
       \sum_{\alpha'}w_{\alpha'}(t) H_{\alpha',\alpha}(t)
\end{equation} 
which can also be derived by complex conjugation of~\eqref{eq:schrocomplexe_z} 
since the hermiticity of~$\hat{H}$ reads~$H_{\alpha,\alpha'}^*=H^{}_{\alpha',\alpha}$. 

The quantum evolution between~$t_i$ and~$t_f$ can therefore be rephrased with a variational principle
based on a functional having the classical form~\eqref{def:actionhamiltonienneB} with a boundary function~$\mathsf{B}(w,z,t)$. Since in this context 
we will not consider transformations of time that depend on the dynamical variables, we can safely
use~$t$ as the integration variable
and work with
\begin{equation}\label{eq:SBwz}
        S_B[w(\cdot),z(\cdot)]
\DEF\!\int_{t_i}^{t_f}\!\!\bigg\{
 \sum_{\alpha}
w_\alpha(t)\,\dot{z}_{\alpha}(t)
-\mathsf{H}(w,z,t)+\frac{\dmat \mathsf{B}}{\dmat t}
\bigg\}\dmat t
\end{equation}
where the complex functions~$t\mapsto{z_\alpha(t)}$ and
$t\mapsto{w_\alpha(t)}$ are considered to be independent one from the other. Together they constitute~$u=(w,z)$ with~$x=t$~($d=1$). 
Thus, all the classical analysis of \S\;\ref{sec:hamvarprinciple} and
\S\;\ref{subsec:finitedof} still holds.
The variations of~$z$ and $w$ are 
constrained by the boundary conditions
\begin{equation}
   \Big[\sum_\alpha w_\alpha\,\delta z_\alpha+\delta \mathsf{B}\Big]^{t_f}_{t_i}=
\Big[\bra{\chi}\big(\delta\ket{\psi}\big)+\delta \mathsf{B}\Big]^{t_f}_{t_i}=0
\end{equation}
where~$\bra{\chi}$ is such that~$w_\alpha=\braket{\chi}{\phi_\alpha}$. 
All the variations~$\delta\ket{\psi}$ of the dynamical variables given by~$\ket{\psi}$ cannot generically
vanish at~$t_i$ and~$t_f$ since there is in general no solution of the Schr\"odinger equation~\eqref{eq:evolutionket}
for an a priori given arbitrary choice of one  initial \emph{and} one final state. Due also to the linear dependence
of the Hamiltonian~$\mathsf{H}$ with respect to~$z$ and~$w$, we cannot express~$p=w$ as 
a function of~$(q,\dot{q})=(z,\dot{z})$  and therefore we
cannot switch to a Lagrangian formulation unless we change completely of strategy and collect the variables~$w$ with the variables~$z$ into the same
 configuration space.

According to Wigner's theorem (see for instance \cite{Simon+08a,Mouchet13b} and their references), a (possibly time-dependent) continuous transformation is represented
by a unitary operator~$\hat{U}$ implemented as follows
\begin{equation}\label{def:Tbraket}
   \T{\phantom{|}}\!\bra{\chi}\DEF\bra{\chi}\hat{U}^*\;;\qquad\T{\ket{\psi}}\DEF\hat{U}\ket{\psi}
\end{equation}
or, with the canonical complex notation,
\begin{equation}
  \sum_{\alpha'}\T{w}_{\alpha'}\bra{\phi_{\alpha'}}\hat{U}\ket{\phi_{\alpha}}=w_\alpha\;;\qquad
  \T{z}_\alpha=\sum_{\alpha'}\bra{\phi_\alpha}\hat{U}\ket{\phi_{\alpha'}}z_{\alpha'}\;.
\end{equation}
By straightforward identification with the complex version of~\eqref{eq:Fpq} with vanishing derivatives
of~$\T{t}$, we have
\begin{subequations}
\begin{eqnarray}
 w_\alpha &=& \frac{\partial\mathsf{F}}{\partial z_\alpha}\;;\\
  \T{z}_\alpha&=&\frac{\partial\mathsf{F}}{\partial \T{w}_\alpha}
\end{eqnarray}  
\end{subequations}  
with the generating function
\begin{equation}
  \mathsf{F}(\T{w},z)
=\sum_{\alpha,\alpha'}\T{w}_\alpha\bra{\phi_\alpha}\hat{U}\ket{\phi_{\alpha'}}z_{\alpha'}
\end{equation}
or, equivalently,
\begin{equation}\label{eq:Fchipsi}
  \mathsf{F}=\T{\phantom{|}}\!\bra{\chi}\hat{U}\ket{\psi}\;.
\end{equation}
For a one-parameter transformation, its generator is a self-adjoint operator~$\hat{G}$, 
possibly time-dependent, defined by
\begin{equation}\label{eq:Uinfinitesimal}
  \hat{U}(\epsilon)=1+\frac{\imat\epsilon}{\hbar}\hat{G}+\Omat(\epsilon^2).
\end{equation}
Then, the Taylor expansion of the generating function~$\mathsf{F}(w,z,t;\epsilon)$ is given by
\begin{equation}
  \mathsf{F}(\T{w},z,t;\epsilon)=\sum_{\alpha}\T{w}_{\alpha}z^{}_{\alpha}
+\frac{\imat\epsilon}{\hbar}\sum_{\alpha,\alpha'}\T{w}_{\alpha}\bra{\phi^{}_\alpha}\hat{G}\ket{\phi^{}_{\alpha'}}
 z^{}_{\alpha'} +\Omat(\epsilon^2)
\end{equation}
from which, by identification with the complexification of~\eqref{eq:dldeF}, we read the ``classical'' generator
\begin{equation}
  \mathsf{G}=\frac{\imat}{\hbar}\sum_{\alpha,\alpha'}w_{\alpha}\bra{\phi_\alpha}\hat{G}\ket{\phi_{\alpha'}}
 z_{\alpha'}=\frac{\imat}{\hbar}\bra{\chi}\hat{G}\ket{\psi}
\end{equation}
of the transformation. 

Now for an invariance we respect the time translations, \eqref{eq:dHdtzero} reads 
\begin{equation}
  0=\frac{\dmat \mathsf{H}}{\dmat t}=\frac{\dmat}{\dmat t}\bra{\chi}\hat{H}\ket{\psi}
=\bra{\chi}\frac{\dmat\hat{H}}{\dmat t}\ket{\psi}
\end{equation}
for any~$\bra{\chi}$ and $\ket{\psi}$, that is we recover
\begin{equation}\label{eq:Hopconstant}
  \frac{\dmat\hat{H}}{\dmat t}=0.
\end{equation}
For an invariance with respect
to a time-independent transformation, \eqref{eq:Gconstant} reads
\begin{equation}
  0=\frac{\dmat \mathsf{G}}{\dmat t}=\frac{\imat}{\hbar}\frac{\dmat}{\dmat t}\bra{\chi}\hat{G}\ket{\psi}
=\frac{\imat}{\hbar}\bra{\chi}\left(\frac{\dmat\hat{G} }{\dmat t}+\frac{\imat}{\hbar}[\hat{H},\hat{G}]\right)\ket{\psi}
\end{equation}
where~$[\ ,\ ]$ denotes the commutator between two operators.
Then we get the identity
\begin{equation}
  \frac{\dmat\hat{G} }{\dmat t}+\frac{\imat}{\hbar}[\hat{H},\hat{G}]=0\;.
\end{equation}
In the Schr\"odinger picture the time-independence of the transformation is equivalent 
to~$\dmat\hat{G}/\dmat t=0$ and therefore the previous identity reduces to
\begin{equation}\label{eq:comHG}
  [\hat{H},\hat{G}]=0
\end{equation}
which is of course the well-known consequence of the invariance of the quantum dynamics
under the transformations generated by~$\hat{G}$.

\subsection{Following Noether's approach}

It is instructive to
 check directly that the results of the previous section can be obtained with more Noether flavour
 by the method of \S\;\ref{subsec:finitedof}.  
In terms of bras and kets we rewrite~\eqref{eq:SBwz} as
\begin{equation}
        S_B[\chi,\psi]
\DEF\!\int_{t_i}^{t_f}\!\!\bigg\{\bra{\chi}\frac{\dmat}{\dmat t}\ket{\psi}+\frac{\imat}{\hbar}\bra{\chi}\hat{H}\ket{\psi}+\frac{\dmat \mathsf{B}}{\dmat t}
\bigg\}\dmat t
\end{equation}
The general expression~\eqref{eq:deltaJdeltau} together with~\eqref{eq:JBJ0} provides
\begin{equation}
         \delta J_B=\bra{\chi}\left(\frac{\dmat}{\dmat t}+\frac{\imat}{\hbar}\hat{H}\right)\ket{\psi}\delta t+\bra{\chi}\big(\delta\ket{\psi}\big)+\frac{\dmat \mathsf{B}}{\dmat t}\delta t+\delta \mathsf{B}\;.
\end{equation}
Moreover, in order to preserve the structure of~$S_B$, we naturally choose the boundary function with the same form
as the Hamiltonian~\eqref{def:Hwz}, that is
\begin{equation}
  \mathsf{B}\DEF\bra{\chi}\hat{B}\ket{\psi}
\end{equation}
for some operator~$\hat{B}$. Then,
the infinitesimal current reads
\begin{equation}
         \delta J_B=\delta J_0+\delta t\frac{\dmat}{\dmat t}\Big(\bra{\chi}\hat{B}\ket{\psi}\Big)
+\bra{\chi}\hat{B}\big(\delta\ket{\psi}\big)+\big(\delta\bra{\chi}\big)\hat{B}\ket{\psi}\;.
\end{equation}
with
\begin{equation}
  \delta J_0=\delta t\bra{\chi}\left(\frac{\dmat}{\dmat t}+\frac{\imat}{\hbar}\hat{H}\right)\ket{\psi}+
\bra{\chi}\big(\delta\ket{\psi}\big)\;.
\end{equation}

The action of~\eqref{eq:Uinfinitesimal} on~$\bra{\chi}$ and on~$\ket{\psi}$ 
leads to
\begin{equation}
\delta\bra{\chi}=-\frac{\imat\epsilon}{\hbar}\bra{\chi}\hat{G}\;;\qquad
\delta\ket{\psi}=\phantom{-}\frac{\imat\epsilon}{\hbar}\hat{G}\ket{\psi}\;.
\end{equation}
Thus,
\begin{equation}\label{eq:deltaJBschro}
  \delta J_B=\delta t\bra{\chi}\left(\frac{\dmat}{\dmat t}+\frac{\imat}{\hbar}\hat{H}\right)\ket{\psi}+
\frac{\imat\epsilon}{\hbar}\bra{\chi}\Big(\hat{G}+[\hat{B},\hat{G}]\Big)\ket{\psi}+\delta t\frac{\dmat}{\dmat t}\Big(\bra{\chi}\hat{B}\ket{\psi}\Big).
\end{equation}

 The transformed  boundary operator is defined to be such 
that
\begin{equation}
  \T{\phantom{|}}\!\bra{\chi}\T{\hat{B}}(\T{t})\T{\ket{\psi}}=\bra{\chi}\hat{B}(t)\ket{\psi}
\end{equation}
for any~$\bra{\chi}$ and~$\ket{\psi}$, that is, by using~\eqref{def:Tbraket},
\begin{equation}
  \T{\hat{B}}(\T{t})=\hat{U}\hat{B}(t)\hat{U}^*\;.
\end{equation}
This identity can also be recovered from~\eqref{eq:Btransforme} with~\eqref{eq:Fchipsi}
by using the complex canonical formalism of the previous section.
The traduction of the invariance is simply~$\T{\hat{B}}(\T{t})={\hat{B}}(\T{t})$ and
then, for an infinitesimal transformation characterized by~$\delta t=\T{t}-t$ and~$\epsilon$,
we get
\begin{equation}
  \delta t\frac{\dmat \hat{B}}{\dmat t}+\frac{\imat\epsilon}{\hbar}[\hat{B},\hat{G}]=0\;.
\end{equation}
If we choose all the operators in the Heisenberg picture, this identity leads to
\begin{equation}\label{eq:dBdtschro}
  \delta t\frac{\dmat \hat{B}}{\dmat t}+\delta t\frac{\imat}{\hbar}[\hat{H},\hat{B}]
+\frac{\imat\epsilon}{\hbar}[\hat{B},\hat{G}]=0\;.
\end{equation}
where all the operators are now considered in the Schr\"odinger 
picture\footnote{By using a label to distinguish the two pictures, for any operator~$\hat{O}$
we have the connection
  \begin{equation} \label{eq:OHUOSU}
        \hat{O}^{\mathrm{\scriptscriptstyle (H)}}(t)
        =\hat{U}^{\mathrm{\scriptscriptstyle (S)}}(t_0,t)\;  
        \hat{O}^{\mathrm{\scriptscriptstyle (S)}}(t)
        \hat{U}^{\mathrm{\scriptscriptstyle (S)}}(t,t_0)
\end{equation}
where~$t_0$ denotes the time where the two pictures coincide 
and~$\hat{U}^{\mathrm{\scriptscriptstyle (S)}}(t,t_0)$ is the evolution operator between~$t_0$ and~$t$
in the Schr\"odinger picture. Therefore we have
\begin{equation}\label{eq:dOdt_qu}
        \frac{\dmat \hat{O}^{\mathrm{\scriptscriptstyle (H)}}(t)}{\dmat t}=
        \frac{\imat}{\hbar}[\hat{H}^{\mathrm{\scriptscriptstyle (H)}}(t),\hat{O}^{\mathrm{\scriptscriptstyle (H)}}(t)]+ \left(\frac{\dmat\hat{O}^{\mathrm{\scriptscriptstyle (S)}}(t) }{\dmat t}\right)^{\mathrm{\scriptscriptstyle (H)}}\;.
\end{equation}
}.
When both~$\ket{\psi}$ and~$\bra{\chi}$ satisfy the Schr\"odinger equation
let us show how the infinitesimal current~\eqref{eq:deltaJBschro} simplifies.
The first  term
in the right-hand side vanishes and the last term is given by
\begin{equation}
  \delta t\frac{\dmat}{\dmat t}\Big(\bra{\chi}\hat{B}\ket{\psi}\Big)=
  \delta t\bra{\chi}\left(\frac{\dmat\hat{B}}{\dmat t}+\frac{\imat}{\hbar}[\hat{H},\hat{B}]\right)\ket{\psi}=-\frac{\imat\epsilon}{\hbar}\bra{\chi}[\hat{B},\hat{G}]\ket{\psi}
\end{equation}
where~\eqref{eq:dBdtschro} has been used for the second equality. Eventually
we obtain 
\begin{equation}
  \delta J_B=\frac{\imat\epsilon}{\hbar}\bra{\chi}\hat{G}\ket{\psi}
\end{equation}
and the conservation law~$\dmat \delta J_B/\dmat t=0$ is exactly equivalent to 
\begin{equation}
  \frac{\dmat }{\dmat t}\Big(\bra{\chi}\hat{G}\ket{\psi}\Big)=0
\end{equation}
from which we already derived~\eqref{eq:Hopconstant} for a model invariant under time-trans\-lations and~\eqref{eq:comHG}
for a model invariant under time-independent transformations.

In passing we note that the Noether constant associated with
the invariance of~$S_B$ under a global change of phase~$\T{\phantom{|}}\!\bra{\chi}=\bra{\chi}\EXP{-\imat\theta}$ together with ~$\T{\ket{\psi}}=\EXP{-\imat\theta}\ket{\psi}$ for any constant~$\theta$
corresponds to~$\hat{G}\propto 1$ and therefore is given by the scalar 
product~$\braket{\chi}{\psi}$ which is indeed conserved by any unitary evolution.

\section{Conclusion}

 Unlike what occurs generically in the Lagrangian context where one remains in the configuration
space,   the Hamiltonian variational principle cannot be formulated 
with keeping fixed all  the dynamical variables  at the boundaries in phase-space. 
Nevertheless, with the use of a boundary function that helps to manage
the issues of boundary conditions,
we have shown how Noether's seminal work \cite{Noether18a}  does cover the Hamiltonian variational
 principle and how the constant generators of the canonical---classical or quantum---transformations are indeed 
the corresponding Noether constants.

\paragraph{Acknowledgement} I am  indebted to Xavier Bekaert for his careful reading of the first proof
of this manuscript and his encouraging comments. I also thank the anonymous referee for her/his comment
that led to footnote~\eqref{fn:addednotev2}.

\bigskip

\end{document}